\documentclass[12pt,preprint]{aastex}
\usepackage{pslatex}
\usepackage{amssymb}

\shorttitle{Magnetic Field and Particle Acceleration}
\shortauthors{Diamond and Malkov}

\newcommand\eg{{\it e.g.,~}}
\newcommand\ie{{\it i.e.,~}}

\newcommand\const{{\rm const\ }}

\begin{document}

\title{Dynamics of Mesoscale Magnetic Field in Diffusive Shock Acceleration}

\author{P.H. Diamond and M.A. Malkov }

\affil{\textit{University of California at San Diego, La Jolla, California
92093-0424, USA }}

\email{e-mail: pdiamond@physics.ucsd.edu; mmalkov@ucsd.edu}

\noindent \textbf{Abstract}

We present a theory for the generation of mesoscale ($kr_{g}\ll 1$,
where $r_{g}$ is the cosmic ray gyroradius) magnetic fields during
diffusive shock acceleration. The decay or modulational instability
of resonantly excited Alfven waves scattering off ambient density
perturbations in the shock environment naturally generates larger
scale fields. For a broad spectrum of perturbations, the physical
mechanism of energy transfer is random refraction, represented by
diffusion of Alfven wave packet in $k-$space. The scattering field
can be produced directly by the decay instability or by the Drury
instability, a hydrodynamic instability driven by the cosmic ray pressure
gradient. This process is of interest to acceleration since it generates
waves of longer wavelength, and so enables the confinement and acceleration
of higher energy particles. This process also limits the intensity
of resonantly generated turbulent magnetic field on $r_{g}$ scales.


\section{Introduction and Overview}

\subsection{Status of the CR Acceleration Problem\label{sub:Status-of-the}}

There is an increasingly popular view that the CR spectrum above the
{}``knee'' (about $10^{15}$eV) is produced by the same accelerator
(or accelerators) as the part below the knee. The latter portion of
the spectrum is better understood. As is usually argued, spectrum
softening at the knee (from $E^{-2.7}$ to $E^{-3.1}$) forces adherents
to the premise of extragalactic origin for CRs above the knee to also
explain why the galactic part of the spectrum terminates at \emph{precisely}
the point where the extragalactic part appears. Therefore it appears
desirable to explain the CR spectrum straight up to its next feature,
the {}``ankle'' (at about $10^{18}$eV) by a \emph{single mechanism}
operating in one accelerator or in a group of similar accelerators.
SNR shocks are now considered to be the most promising candidate for
that purpose. Recent discussion of available suggestions that seek
to accelerate cosmic rays (CRs) in supernova remnants (SNR) to energies
beyond the {}``knee'' can be found in \markcite{Pariz04,Hillas05}{Parizot} {et~al.} (2004); {Hillas} (2005).

Of course, SNRs are almost certainly responsible for the CR (at least
electron) acceleration below the knee, which is documented in several
ways (see \markcite{aharbook}{Aharonian}, 2004 for a comprehensive review of detection
techniques and physical processes). In particular, there is an evidence
(\markcite{Koyama95,Tanimori98,Allen01}{Koyama} {et~al.} 1995; {Tanimori} {et~al.} 1998; {Allen}, {Petre}, \& {Gotthelf} 2001; in the form of both synchrotron
and inverse Compton radiation) that electrons of energies up to $100TeV$
are accelerated in the supernova shock waves. Recently, accelerated
electrons with record energies up to at least $450$TeV have been
discovered in a young shell-type supernova remnant G12.8-0.0 \markcite{Brogan05}({Brogan} {et~al.} 2005).
Interestingly enough, this source was previously detected in the $\gamma $-ray
band by the HESS Cerenkov telescope \markcite{AharSci05}({Aharonian} {et~al.} 2005).

By association, the electron acceleration mechanism should be also
responsible for accelerating the main CR component, namely the protons.
Indeed, at ultrarelativistic energies particle dynamics is determined
by particle momentum (\ie rigidity) regardless of mass. However,
the acceleration of protons in supernova remnant (SNR) shocks has
not been confirmed conclusively. The only SNR where a signature of
accelerated protons was claimed to be observed is the RXJ 1713.7-3946
\markcite{Enomoto02}({Enomoto} {et~al.} 2002). This source was first detected in the $\gamma $-
ray band by the CANGAROO team \markcite{Muraishi00}({Muraishi} {et~al.} 2000), and later confirmed
by HESS \markcite{AharNat04}({Aharonian} {et~al.} 2004). However, the claim made by the CANGAROO
team \markcite{Enomoto02}({Enomoto} {et~al.} 2002) became a subject of significant controversy
\markcite{Pohl02,ButtNat02}({Reimer} \& {Pohl} 2002; {Butt} {et~al.} 2002), primarily on the grounds that explanation
of the data would require a break in the energy spectrum of accelerated
protons, while the \emph{standard} diffusive shock acceleration (DSA)
mechanism does not predict the necessary break. A possible resolution
of this controversy, based on nonlinear effects in the diffusive acceleration
mechanism, along with the generation of a break due to modification
of particle confinement in molecular clouds surrounding the remnants
(via neutral particle effect on Alfven waves) was recently suggested
by \markcite{MDS05}{Malkov}, {Diamond}, \& {Sagdeev} (2005). The latest high resolution observations with HESS
\markcite{Ahar06RXJ}({Aharonian} {et~al.} 2006) are also consistent with the presence of a break
in the spectrum in the TeV energy range, but only future, more accurate
data will be able to distinguish between the different possible functional
forms of the high energy spectrum predicted by different models. The
simplest (and by far the most popular) approach, based on a power-law
spectrum with an exponential cut-off, produces almost as good a fit
to the current data as the spectrum with a break \markcite{Ahar06RXJ}({Aharonian} {et~al.} 2006)
does. However, \emph{the spectral break is a distinguishing feature},
that reveals valueable information about the physical processes responsible
for its formation and about acceleration in general. These processes
include, but are not limited to, quasi-abrupt changes in 

\begin{enumerate}
\item the dynamics of waves that confine particles (\markcite{mdj02,DM04,MDS05}{Malkov}, {Diamond}, \& {Jones}, 2002; {Diamond} \& {Malkov}, 2004; {Malkov} {et~al.}, 2005) 
\item the overall acceleration regime \markcite{Drury03aph}({Drury}, {van der Swaluw}, \&  {Carroll} 2003), and 
\item the particle confinement regime \markcite{MD06}({Malkov} \& {Diamond} 2006). 
\end{enumerate}
Further broad band studies of RXJ 1713.7-3946 are also needed to fully
resolve the dilemma of hadronic versus leptonic origin of the TeV
emission. Recent TeV observations with HESS \markcite{AharNat04}({Aharonian} {et~al.} 2004) allowed
authors to suggest that both leptonic and hadronic components are
being accelerated in this object. Furthermore, the latest observations
with the same instrument \markcite{Ahar06RXJ}({Aharonian} {et~al.} 2006) seem to indicate a decline
in the proton spectrum at energies $\sim 100$TeV if the observed
spectrum can be interpreted as a by-product of the interaction of
accelerated protons with the ambient gas. From a theoretical standpoint,
a very interesting aspect of this spectrum is that it is \emph{significantly}
softer than one would expect from a strong shock in a nonlinear (spectral
index$\simeq 1.5$) or even in a linear regime (spectral index$=2.0$).
Should future measurements confirm that it must be interpreted as
a piece of a broken power-law spectrum \markcite{MDS05}({Malkov} {et~al.} 2005), rather than
its cut-off \markcite{Berezh06RXJ}({Berezhko} \& {Voelk} 2006), a new interpretation requiring new
DSA physics, along the lines of (1-3) above will be necessary. Note
that current numerical models (\eg\markcite{Berezh06RXJ}{Berezhko} \& {Voelk}, 2006) do not
include the wave-particle interaction effects necessarily involved
in (1) and (3). Unfortunately, even the high quality TeV data of \markcite{Ahar06RXJ}({Aharonian} {et~al.} 2006)
degrades statistically at photon energies above $10$ TeV, where the
spectrum begins to decline. Nevertheless, the spectrum appears more
like a broken power-law at $10$ TeV rather than a power-law with
an exponential cut-off. The latter is more consistent with the time-dependent
acceleration or energy loss scenario (see, \eg Fig.12 in \markcite{Ahar06RXJ}{Aharonian} {et~al.}, 2006,
and Fig.8 in \markcite{Berezh06RXJ}{Berezhko} \& {Voelk}, 2006) implemented in most numerical
models. 

Interestingly, even observations of the background CRs pose similar
problems to the standard acceleration theory. The most notorious one
is perhaps the overall CR spectrum itself which, as discussed earlier,
is too steep in its high energy part (\ie above the {}``knee'')
to be explained straightforwardly by standard acceleration theory.
Leaving aside energy dependent propagation of CRs as well as their
reacceleration (which may also significantly influence their spectrum,
as noted by \markcite{PtusCOSP97}{Ptuskin} {et~al.}, 1997), one can distinguish two problems.
The first problem is the physical origin of the break ({}``knee'')
and the second problem is the very stringent requirements on the acceleration
parameters, such as the turbulent component of the magnetic field,
in order to reach to the very high {}``ankle'' energy of $\sim 10^{18}eV$.

\subsection{Approaches to Enhanced DSA\label{sec:Possible-approaches}}

The important quantities that regulate diffusive shock acceleration
(DSA) are the strength and spectral distribution of the turbulent
magnetic field, $\delta B$. The magnetic turbulence confines accelerated
particles to the shock front by pitch-angle scattering and is believed
to be produced by the particles themselves, via CR-Alfven wave resonance
when CRs stream ahead of the shock. Within standard quasi-linear theory
(which is strictly valid only for $\delta B\ll B_{0}$), pitch angle
scattering of relativistic particles by the waves proceeds at the
rate 

\begin{equation}
\nu \sim \Omega \frac{mc}{p}\left(\frac{\delta B}{B_{0}}\right)^{2}\label{eq:nu}\end{equation}
where $\Omega $ and $p$ are the (nonrelativistic) gyrofrequency
and momentum. Resonance requires $kr_{g}=\const $. Particle self-confinement
along the field is diffusive, and the diffusivity is $\kappa \sim c^{2}/\nu $.
The acceleration time scale can be estimated as $\tau _{acc}\sim \kappa /u_{s}^{2}$,
where $u_{s}$ is the shock speed. The fluctuating part of the field
is usually assumed to saturate at the level of the ambient field,
$\delta B\sim B_{0}$, which thus produces pitch-angle scattering
at the rate of $\Omega $ (the gyrofrequency), which limits the particle
mean free path (m.f.p.) along the field to a distance of the order
of gyroradius. This constitutes the so called {}``Bohm diffusion
limit''. Under these circumstances, the mean field $B_{0}$ sets
the acceleration rate and the maximum particle energy. The latter
can be expressed through the work done by the induced electric field
$\left(u_{s}/c\right)B_{0}$ on the particles while they are carried
with the shock at the speed $u_{s}$ over the length scale of the
shock radius, $R_{s}$. Thus, the maximum energy is $E_{max}\sim (e/c)u_{s}B_{0}R_{s}$.
Thus, one arrives at the maximum CR energy accelerated in a typical
SNR shock of about $10^{15}$eV, which is close to the {}``knee''
but is three orders of magnitude below the {}``ankle''. Before discussing
any approaches to enhanced, beyond-the-{}``knee'', acceleration,
it is important to note that, due to the resonance condition $kp=const$,
confinement of higher energy particles requires that longer waves
must be excited. Put another way, \emph{inverse scattering or transfer
of Alfven wave energy excited at} $kr_{g}\sim 1$ \emph{to longer
scales is clearly beneficial to confinement and acceleration}.

\begin{figure}
\epsscale{.50}
\plotone{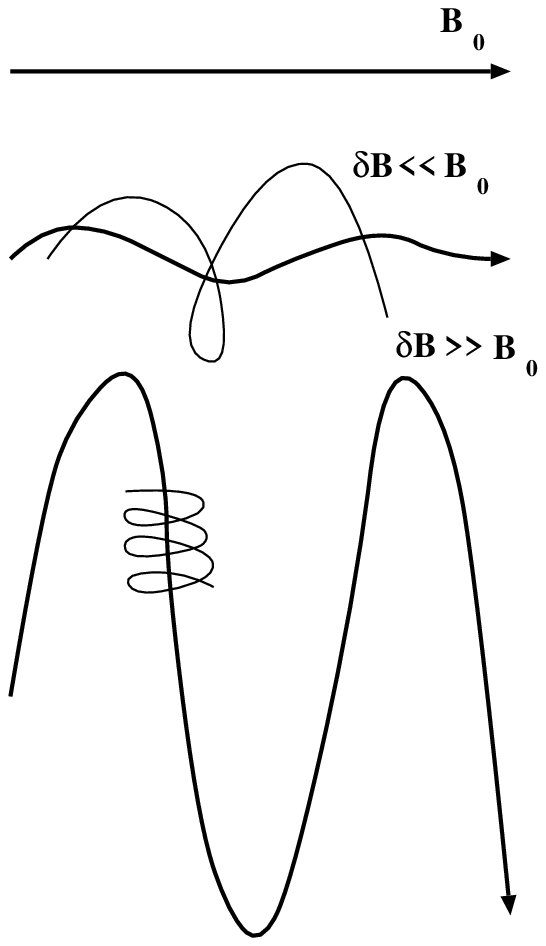}

\caption{Weak and strong types of magnetic fluctuations with superimposed
particle trajectories.\label{cap:Weak-and-strong} }
\end{figure}

In order to reach the energy of the {}``ankle'' several suggestions
have been made. One approach is to invoke the generation of a fluctuating
field component $\delta B$ significantly in excess of the unperturbed
field $B_{0}$ \markcite{LucBell00}({Lucek} \& {Bell} 2000). Physically, such generation is
deemed possible since the free energy source is the pressure gradient
of accelerated particles, which may reach a significant fraction of
the shock ram energy. Specifically, a free energy limit on the wave
energy density $\left(\delta B/B_{0}\right)^{2}$ may be related to
the partial pressure $P_{c}$ of CRs that resonantly drive the waves
by the relation \markcite{McKVlk82}({MacKenzie} \& {Voelk} 1982)

\begin{equation}
\left(\delta B/B_{0}\right)^{2}\sim M_{A}P_{c}/\rho u_{s}^{2}.\label{delB}\end{equation}
Here $M_{A}=u_{s}/V_{A}\gg 1$ is the Alfven Mach number and $\rho u_{s}^{2}$
is the shock ram pressure. Of course, when $\delta B/B_{0}$ exceeds
unity, particle dynamics, and thus the particle confinement and acceleration
rates depart radically from the quasi-linear picture that underpins
the usual DSA theory and modeling. The simple case of a monochromatic
wave with arbitrary $\delta B/B_{0}$, in which particle dynamics
are exactly integrable \markcite{Lutom66}({Lutomirski} \& {Sudan} 1966), provides an important clue
to the general case (\eg \markcite{m98}{Malkov}, 1998). A critical parameter is
$kr_{g}^{*}$ where $k$ is the wave number and $r_{g}^{*}$ \emph{is
the particle Larmor radius calculated with the perturbed} $\delta B\gg B_{0}$
\emph{field, rather than} $B_{0}$, Fig.\ref{cap:Weak-and-strong}.
Particles with $kr_{g}^{*}\la 1$ perceive a strong local field which
is perpendicular to $\mathbf{B}_{0}$ and therefore their confinement
in the $\mathbf{B}_{0}$ direction is good. Particles with $kr_{g}^{*}>1$
perceive \emph{only the averaged, rather than the local field}, which
is weak. The $\delta B$ component (even if it is large) exerts only
a (rapidly) \emph{oscillating} force on these particles, which thus
can escape along $B_{0}$. Put more mathematically, for $kr_{g}^{*}\gg 1$,
the particle response is the Boltzmann response. For the resonantly
driven waves $kr_{g}\sim 1$, and so if the waves grow nonlinearly
until $r_{g}^{*}\ll r_{g}$, the particles that initially destabilize
the waves are trapped by the wave, thus saturating the instability
in the wave band corresponding to their energy \markcite{LucBell00}({Lucek} \& {Bell} 2000).
This is analogous to the saturation of the beam-plasma instability
for the so called hydrodynamic regime, for which the wave stops growing
when its amplitude is sufficient to trap the beam particles. The numerical
studies by \markcite{LucBell00}{Lucek} \& {Bell} (2000) showed that at least for the case of
an MHD background plasma and \emph{rather narrow wave (and particle
energy) band}, the amplitude of the principal mode can reach a few
times that of the background field. Moreover, \markcite{BelLuc01}{Bell} \& {Lucek} (2001) argue
that in the case of efficient acceleration, field amplification may
be even stronger, reaching a $mG$ level from the background of a
few $\mu G$ ISM field, thus providing acceleration of protons up
to $10^{17}$ eV in SNRs. In view of the above discussion, however,
we would like to emphasize that the simulations of Bell and Lucek
are intrinsically limited precisely because they are narrowband. Thus,
wave-particle interaction is quite restricted, and the simulations
do not allow any interaction of Alfven waves with other types of fluctuations
which, in reality, are likely to be present in a shock environment.
Moreover, Bell and Lucek did not address the question of how to compute
the particle scattering rate for $\delta B\gg B_{0}$. Certainly,
it is not correct to simply plug such a large amplitude into the usual
quasi-linear diffusion coefficient, equation (\ref{eq:nu}).

Recently, \markcite{PtusZir03}{Ptuskin} \& {Zirakashvili} (2003) approached this problem from a different
perspective. They considered a Kolmogorov-type turbulent cascade to
small scales, assuming the waves are generated by efficiently accelerated
particles on the long-wave part of the spectrum. The question of why
such a Kolmogorov-like cascade model is relevant to a shock environment
was not addressed. They obtained a maximum particle energy similar
to the result of \markcite{BelLuc01}{Bell} \& {Lucek} (2001). Yet another consequence of high
magnetic field fluctuation levels, which is based on the change of
the expansion regime of the SNR shocks, is discussed by \markcite{Drury03aph}{Drury} {et~al.} (2003).
An important question that still remains, is \emph{how realistic the
saturation level given by} equation (\ref{delB}) \emph{is}? Earlier
studies by \markcite{Voelk84}{Voelk}, {Drury}, \& {McKenzie} (1984) and \markcite{AchtBland86}{Achterberg} \& {Blandford} (1986) suggest that
due to particle trapping the instability saturates at levels $\delta B\sim B_{0}$.
 For Alfven \emph{turbulence}, consideration of random parallel scattering
(\ie turbulent mirroring) \markcite{Acht81a,AchtBland86}({Achterberg} 1981; {Achterberg} \& {Blandford} 1986) implies a
similar saturation level. Thus, the questions of the saturation level
and the mechanism of confinement of high energy particles remain unanswered.

Motivated by the above issues and by the problems in describing the
observed high-energy spectra as outlined in the Introduction, \markcite{MD06}{Malkov} \& {Diamond} (2006)
suggested a faster-than-Bohm rate acceleration which is also intimately
related to the knee and other spectral break phenomenon. The mechanism
does not require $\delta B/B_{0}\gg1$ magnetic field fluctuations,
since particles gain energy by bouncing between the scatterers convected
with the gradually converging upstream flow in a nonlinearly modified
shock precursor. This is different than the bouncing between the upstream
and downstream scattering centers, which is usually assumed in DSA
models. Note that \emph{in the latter case the acceleration rate decreases
with energy, on account of increasing particle mean free path} (particle
diffusion length at the shock). This generic deficiency of the DSA
is not relevant to the recently proposed mechanism, since the increase
in m.f.p. is exactly compensated by the increased flow-induced compression
of the scattering centers between which the particle bounces in the
precursor flow. The maximum momentum is estimated to be

\begin{equation}
p_{max}\sim \frac{c}{u_{sh}}\frac{L}{L_{p}}p_{*}\label{eq:pmaxfin}\end{equation}
where $L/L_{p}$ is the ratio of the distance between the scattering
centers (weak shocks) to the precursor length (roughly estimated to
be $\la 10$) and $p_{*}$ is the maximum momentum achieved during
the standard phase of the DSA, which becomes the break point or {}``knee''
of the final spectrum. The spectral index between $p_{*}$ and $p_{max}$
is steeper than the {}``standard'' $p^{-4}$ and its slope depends
on details of particle interaction with scatterers. The acceleration
time is 

\begin{equation}
\tau _{acc}\left(p_{max}\right)\sim \tau _{NL}\left(p_{*}\right)\ln \frac{p_{max}}{p_{*}}\label{eq:tauaccfin}\end{equation}
 where $\tau _{NL}\left(p\right)\simeq 4\kappa _{B}\left(p\right)/u_{sh}^{2}$
(with the Bohm diffusion $\kappa _{B}$) is the nonlinear acceleration
time \markcite{MDru01}({Malkov} \& {Drury} 2001) which only slightly differs from the upstream
contribution to the standard linear acceleration time.

\subsection{Inverse Cascade and Enhanced DSA\label{sub:Inverse-Cascade-and}}

The theory of enhanced acceleration via bouncing between scatterers
undergoing compression in the precursor flow described above does
not specifically address magnetic field effects. Indeed, one is naturally
motivated to ask whether it is, in fact, \emph{possible} to achieve
$\delta B\gg B_{0}$ in a turbulent environment where many types of
different wave interactions are possible. If such high levels are
not achieved, one then must confront the issue of how to confine high
energy particles to the shock. To this end, \markcite{DM04}{Diamond} \& {Malkov} (2004) previously
suggested an acceleration scenario in which the magnetic field may
interact strongly with the shock as a result of the acceleration itself,
which, in turn, may in fact be strongly enhanced. The mechanism of
such enhancement is based on the transfer of magnetic energy to longer
scales via wave-wave interaction, which we call \emph{inverse cascade}
for short, even though specific mechanisms of such transfer may differ
from what is usually understood as a local, self-similar cascade in
MHD turbulence. This transfer is limited only by an outer scale $L_{out}$-
likely the shock precursor size $\kappa (p_{max})/u_{s}\sim r_{g}(p_{max})c/u_{s}\gg r_{g}(p_{max})$.
The conceptual picture of this process is simple. There are two populations
of fluctuations naturally native to a shock environment. These are 

\begin{enumerate}
\item Alfven waves, resonantly generated at small scales, \ie $kr_{g}\sim 1$ 
\item acoustic waves and density fluctuations, at $kL_{out}>1$ but not
$\gg 1$. Acoustic modes can be generated by many processes. The density
perturbation field naturally refracts the Alfven wave field, \ie
\end{enumerate}
\[
\frac{dk}{dt}=-\frac{\partial }{\partial x}\omega =-\frac{\partial }{\partial x}kV_{A}\simeq \frac{kV_{A}}{2}\frac{\partial }{\partial x}\frac{\widetilde{\rho }}{\rho _{0}}\]
For a random array of scatterers, $k$ evolves diffusively, so diffusion
of the Alfven wave spectrum results, with 

\[
D_{k}\simeq \frac{k^{2}V_{A}^{2}}{4}\sum _{q}\left|\frac{\widetilde{\rho }_{q}}{\rho _{0}}\right|^{2}q^{2}\tau _{cq}\]
This diffusion naturally spreads the Alfven wave population in $k$,
and so both prevents the development of a narrow spatial band or beam
with $\delta B/B_{0}\gg 1$. This scattering also generates longer
waves, which can in turn confine higher energy particles. Here, the
Alfven wave population density flux is simply 

\[
\Gamma _{k}=-D_{k}\frac{\partial N}{\partial k}\]
where $N$ is the Alfven wave population density (\ie wave action
density $N=\epsilon /\omega $). Since $\partial N/\partial k>0$
for $kr_{g}<1$ (\ie Alfven waves are excited at $kr_{g}\la 1$),
the flux is \emph{toward} larger scales and lower $k$'s. It is important
to note that this process is non-local in wavenumber, and thus \emph{not}
a true cascade in the traditional sense, and also is not a 'dynamo',
but rather a \emph{re-distribution} of magnetic energy among different
scales. This approach is in distinct contrast to the models of \markcite{BelLuc01}{Bell} \& {Lucek} (2001)
and \markcite{PtusZir03}{Ptuskin} \& {Zirakashvili} (2003) discussed above, which work with magnetic fields
on scale lengths of order of Larmor radius $r_{g}(p_{max})$ of the
highest energy particles, and even smaller. The advantage of an 'inverse
cascade' for acceleration is that longer waves confine higher energy
particles and that the turbulent field at the outer scale $\delta B(L_{out})\equiv B_{rms}$
(generated by the flux in wave number space) which necessarily must
have long autocorrelation time can likely be regarded as an {}``ambient
field'', so far as accelerated particles of all energies are concerned.
If $B_{rms}\gg B_{0}$, then the acceleration can be enhanced by a
factor $B_{rms}/B_{0}$. Note that the resonance field $\delta B(r_{g})$
remains smaller than $B_{rms}$, so that standard arguments about
Bohm diffusion apply. Since the fluctuations around the new 'background'
field $B_{rms}$ remain relatively weak, they are not likely to dissipate
so rapidly via nonlinear processes, such as induced scattering on
thermal protons, as it is to be expected in the case $\delta B\ga B_{0}$
when the transfer of energy from resonant fluctuations $\delta B$
to the large scale $B_{rms}$ is not taken into account. In particular,
such nonlinear processes were not included in the enhanced acceleration
models of \markcite{LucBell00,BelLuc01}{Lucek} \& {Bell} (2000); {Bell} \& {Lucek} (2001). These can reduce the resonant
fluctuation field significantly.

As should be clear from the discussion above, an adequate description
of the acceleration mechanism must treat \emph{both} particle and
wave dynamics on an equal footing. In fact, the situation is even
more difficult, since the acceleration process turns out to be so
efficient that the pressure of accelerated particles markedly modifies
the structure of the shock by both the overall shock compression and
the flow profile. Historically, these three aspects of the mechanism
have been considered in isolation. First, a test particle theory was
formulated, in which wave generation was only tacitly implied in the
prescribed particle diffusivity, and the backreaction of accelerated
particles on the shock structure was neglected. The latter was first
included in the framework of the so called two-fluid model, where
the accelerated particles contribute to the energy and momentum fluxes
across the shock, but were assumed to be massless \markcite{axf77,Drury81}({Axford}, {Leer}, \& {Skadron} 1977; {Drury} \& {Voelk} 1981).
Later, kinetic models were developed, both numerically and analytically.
Early theories that include wave generation selfconsistently with
the acceleration process, in turn neglect particle back-reaction on
the shock \markcite{Bell78}({Bell} 1978). The importance of a self-consistent treatment
of the nonlinear modification of the shock structure and wave propagation
has been demonstrated by \markcite{mdj02}{Malkov} {et~al.} (2002). That this three-way coupling
(between particles, waves and flow) is indeed strong can be understood
by considering compression of particle-generated Alfven turbulence
in a nonlinearly modified (converging) plasma flow ahead of the shock.
Since the wave number of the Alfven waves (which are almost frozen
into the flow, $V_{A}\ll V_{shock}$) is increasing because of compression,
particles with highest energies can no longer interact resonantly
with the waves and so simply leave the system along the field lines.
This limits the acceleration rather naturally as soon as particle
pressure reaches a nonlinear level (comparable to the flow ram pressure),
sufficient to modify the flow and cause a significant wave compression.
Generation of longer wavelength, larger scale waves, as described
above to initiate this process is clearly desireable. We began to
study this process in \markcite{DM04}({Diamond} \& {Malkov} 2004) and we continue this study in
this paper. In the spirit of the discussion of this paragraph, we
treat both wave and particle kinetics on an equal footing in the analysis
presented here. 

The remainder of this paper is organized as follows. Section \ref{sec:Accelerated-Particles-plasma}
presents the overall structure of the theory in the context of a discussion
of accelerated particles, plasma flow and waves near the shock front.
Section \ref{sec:Wave-dynamics-in} discusses the dynamics of wave
interactions in the shock precursor. It is divided into three subsections,
dealing with Alfven wave turbulence, acoustic wave turbulence, and
the dynamics of Alfvenic-acoustic coupling. Section \ref{sec:Calculation-of-Wave}
presents the theory of induced diffusion of Alfven wave quanta. Section
\ref{sec:Mechanisms-of-Magnetic} deals with mechanisms of energy
transfer to larger scales. Section \ref{sec:Conclusions-and-Discussion}
presents conclusions and a discussion.

\section{Accelerated particles, plasma flow and waves near the shock front\label{sec:Accelerated-Particles-plasma}}

The transport and acceleration of high energy particles (CRs) near
a CR modified shock is usually described by a diffusion-convection
equation. It is convenient to use a distribution function $f(p)$
normalized to $p^{2}dp$.

\begin{equation}
\frac{\partial f}{\partial t}+U\frac{\partial f}{\partial x}-\frac{\partial }{\partial x}\kappa \frac{\partial f}{\partial x}=\frac{1}{3}\frac{\partial U}{\partial x}p\frac{\partial f}{\partial p}\label{dc1}\end{equation}
 Here $x$ is directed along the shock normal which, for simplicity,
is assumed to be the direction of the ambient magnetic field. The
two quantities that control the acceleration process are the flow
profile $U(x)$ and the particle diffusivity $\kappa (x,p)$. The
first one is coupled to the particle distribution $f$ through the
equations of mass and momentum conservation

\begin{eqnarray}
\frac{\partial }{\partial t}\rho +\frac{\partial }{\partial x}\rho U=0 &  & \label{mas:c}\\
\frac{\partial }{\partial t}\rho U+\frac{\partial }{\partial x}\left(\rho U^{2}+P_{\mathrm{c}}+P_{\mathrm{g}}\right)=0 &  & \label{mom:c}
\end{eqnarray}
 where

\begin{equation}
P_{{c}}(x)=\frac{4\pi }{3}mc^{2}\int _{p_{inj}}^{\infty }\frac{p^{4}dp}{\sqrt{p^{2}+1}}f(p,x)\label{Pc}\end{equation}
 is the pressure of the CR gas and $P_{\mathrm{g}}$ is the thermal
gas pressure. The lower boundary $p_{inj}$ in momentum space, which
separates CRs from the thermal plasma, enters the equations through
the magnitude of $f$ at $p=p_{inj}$. This value specifies the injection
rate of thermal plasma into the acceleration process. The particle
momentum $p$ is normalized to $mc$. Note that the two-fluid model
can be derived from the system of eqs. (\ref{dc1}-\ref{Pc}) by taking
the energy moment of equation (\ref{dc1}). The spatial diffusivity
$\kappa $, induced by pitch angle scattering, prevents particle streaming
away from the shock, and thus facilitates acceleration by ensuring
the particle completes several shock crossings.

Both two fluid and kinetic treatment of the system (\ref{dc1}-\ref{Pc})
indicate a marked departure from test particle theory. Perhaps the
most striking result of the nonlinear treatment is the bifurcation
of shock structure (in particular the shock compression ratio) in
the parameter space formed by the injection rate, shock Mach number,
and maximum particle momentum \markcite{mdv00}({Malkov}, {Diamond}, \& {V{\"o}lk} 2000). In particular, for sufficiently
strong shocks and high particle energies the transition from the test
particle (unmodified) shock solution to the strongly modified, efficiently
accelerating shock solution is \emph{not} gradual. It has been hypothesized
that in the critical range of parameters other physical processes
must play a crucial role. This include plasma heating and a modified
particle confinement regime, both of which are intimately related
to the wave and turbulence dynamics. We consider this in the next
section.

\section{Dynamics of Wave Interactions in the CR Shock Precursor\label{sec:Wave-dynamics-in} }

The mechanism of transfer or scattering of Alfven wave energy to larger
scales is rather different from that associated with the conventional
picture of a turbulent MHD dynamo. Most notably, it is a \emph{redistribution}
of wave energy from $kr_{g}\sim 1$ to $kr_{g}<1$, where it can consequently
confine higher energy particles, since $kp/m=\Omega $ at resonance.
This process is \emph{not} one of mean field generation or magnetic
flux amplification, though the transfer to the large scale \emph{can}
generate an \emph{apparent} $B_{rms}$ on those scales. Here, magnetic
fluctuations are produced via the familiar process of cyclotron resonance
of cosmic rays. The energy \emph{transfer} mechanism under study is
simply a \emph{decay or modulational instability}, which is a nonlocal
transfer of Alfven wave energy to larger scales, via scattering off
density perturbations in the shock precursor. In contrast to an inverse
cascade, this transfer is \emph{nonlocal} in scale. Also, we note
that the shock precursor is itself linearly unstable to acoustic perturbations.
This mode is called the Drury instability, and assures a well populated
scattering field of density perturbations off which Alfven waves scatter.
Given the plausible assumption that the large scale scatterer field
is stochastically distributed (\ie consists of randomly phased acoustic
waves), the effect of the decay process on the Alfven wave spectrum
is to produce a random walk in $k$, via random refraction.

\begin{figure}
\epsscale{1.0}
\plotone{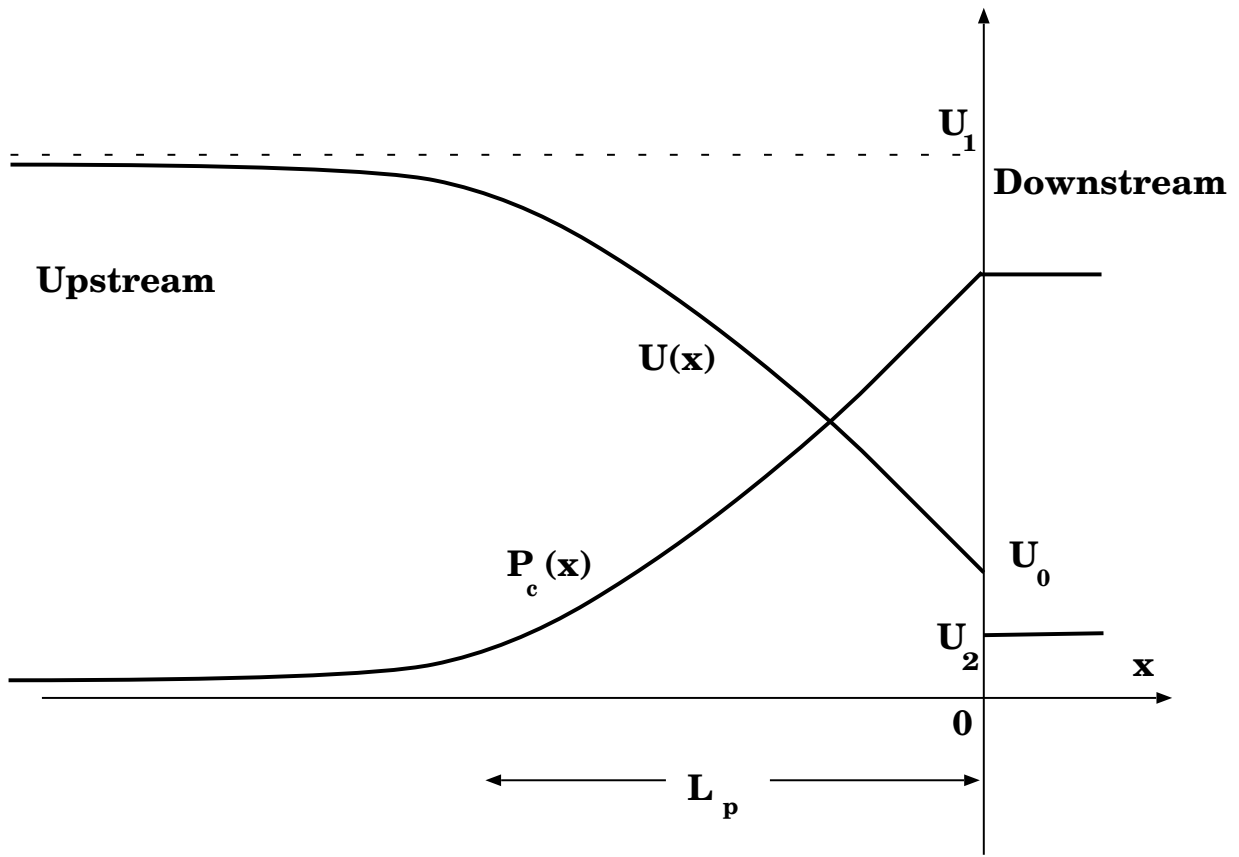}

\caption{The structure of a CR modified shock, with the flow profile $U\left(x\right)$
and CR pressure distribution $P_{c}\left(x\right)$. The sub-shock
is at $x=0$ and the CR precursor of the length $L_{p}$ is formed
upstream, $x<0$.\label{cap:The-structure-of}}
\end{figure}

The spatial structure of an efficiently accelerating shock, \ie 
a shock that transforms a significant part of its energy into accelerated
particles, is very different from that of an ordinary shock. The most
extended part of the shock structure consists of a \emph{precursor}
formed by the cloud of accelerated CRs diffusing ahead of the shock.
If the CR diffusivity $\kappa (p)$ depends linearly on particle momentum
$p$ (as in the Bohm diffusion case), then, at least well inside the
precursor, the velocity profile $U(x)$ is approximately \emph{linear}
in $x$ \markcite{m97a}({Malkov} 1997) where $x$ points downstream antiparallel to
the shock normal, Fig.\ref{cap:The-structure-of}. Ahead of the shock
precursor, the flow velocity tends to its upstream value, $U_{1}$,
while on the downstream side it undergoes a conventional plasma shock
transition to its downstream value $U_{2}$ (all velocities are taken
in the shock frame). This extended CR precursor (of the size $L_{CR}\sim \kappa (p_{max})/U_{1}$)
is the place where we expect turbulence to be generated by the CR
streaming instability and where it couples to longer wavelengths.
In particular, the density fluctuations in the precursor are what
refracts the Alfven waves (produced by the same energetic particles
which form the precursor) and scatter them to larger scales.

\subsection{Alfven Wave Turbulence}

The growth rate of the ion-cyclotron instability is positive for the
Alfven waves traveling in the CR streaming direction \ie upstream,
and it is negative for oppositely propagating waves. The wave kinetic
equation for both types of waves can be written in the form

\begin{equation}
\frac{\partial N^{\pm }}{\partial t}+\frac{\partial \omega ^{\pm }}{\partial k}\frac{\partial N^{\pm }}{\partial x}-\frac{\partial \omega ^{\pm }}{\partial x}\frac{\partial N^{\pm }}{\partial k}=\gamma _{k}^{\pm }N^{\pm }+C^{\pm }\left\{ N^{+},N^{-}\right\} \label{wke}\end{equation}
 Here $N^{\pm }\left(k,x,t\right)$ denote the population of quanta
propagating in the upstream and downstream directions, respectively.
Also, $\omega ^{\pm }$ are their Alfven wave frequencies, $\omega ^{\pm }=kU\pm kV_{A}$,
where $V_{A}$ is the Alfven velocity. The linear growth ($\gamma ^{+}$)
and damping ($\gamma ^{-}$) rates are nonzero only in the resonant
part of the spectrum, for which $kr_{g}(p_{max})\ge 1$, \ie $\gamma ^{\pm }=\gamma ^{\pm }\left(k\right)$.
In the most general case, the last term on the r.h.s. of equation
(\ref{wke}) represents nonlinear interaction of different types of
quanta $N^{+}$ and $N^{-}$, and also compressible MHD self-interaction
of each population via steepening. As seen from these equations, the
coefficients in the wave transport part of this equation (\ie  l.h.s.)
depend on the parameters of the medium through $U$ and $V_{A}$,
which in turn, may be perturbed by slow, large scale fluctuations.
This usually results in parametric or modulational phenomena \markcite{SagdGal69}({Sagdeev} \& {Galeev} 1969).
We will concentrate on the acoustic type perturbations (which may
be induced by Drury instability), so that we can write for the density
$\rho $ and velocity~$U$

\[
\rho =\rho _{0}+\widetilde{\rho };\; U=U_{0}+\widetilde{U}\]
 The variation of the Alfven velocity $\widetilde{V}_{A}=V_{A}-V_{A0}$
is then

\[
\widetilde{V}_{A}\simeq -\frac{1}{2}V_{A}\frac{\widetilde{\rho }}{\rho _{0}}.\]
 For simplicity, we assume that the plasma $\beta <1$, which is valid
upstream of the subshock but not downstream. Since we are primarily
interested in the upstream, shock precursor turbulence, this assumption
is at least reasonable, though one should address the $\beta \sim 1$
case as well. We will consider this in a future publication. In the
downstream medium of a highly superalfvenic shock, clearly $\beta \gg 1$.
We neglect the variation of $U$ compared to that of $V_{A}$ in equation
(\ref{wke}). Note that fluctuations in $U$ merely diffuse the location
of the Alfven wave population in the precursor flow field. The perturbations
of $V_{A}$ in turn induce perturbations of $N_{k}^{\pm }$, so we
can write

\[
N^{\pm }=\left\langle N^{\pm }\right\rangle +\widetilde{N}^{\pm }\]
where $\left\langle N^{\pm }\left(k,x,t\right)\right\rangle $ is
the quanitty of interest, namely the mean wave population. This is
obtained via quasi-linear theory, applied to to the wave kinetic equation
in the same way it is usually applied to the particle kinetic equation.
Given that our goal is to obtain an evolution equation for the average
number of Alfven quanta $\left\langle N^{\pm }\right\rangle $, averaging
equation (\ref{wke:av}) then yields

\begin{equation}
\frac{\partial }{\partial t}\left\langle N^{\pm }\right\rangle +\left(U\pm V_{A}\right)\frac{\partial }{\partial x}\left\langle N^{\pm }\right\rangle -kU_{x}\frac{\partial }{\partial k}\left\langle N^{\pm }\right\rangle +\frac{\partial }{\partial k}\left\langle kV_{A}\frac{\widetilde{\rho }_{x}}{\rho _{0}}\widetilde{N}^{\pm }\right\rangle =\gamma _{k}^{\pm }\left\langle N^{\pm }\right\rangle +\left\langle C\left(N^{\pm }\right)\right\rangle \label{wke:av}\end{equation}
 Here the subscript $x$ stands for the $x$-derivatives. In order
to calculate the correlator $\left\langle \frac{\widetilde{\rho }_{x}}{\rho _{0}}\widetilde{N}^{\pm }\right\rangle $
in equation (\ref{wke:av}) via quasi-linear closure, one must first
determine the coherent response $\widetilde{N}^{\pm }\left(k,x,t\right)$
to the density perturbation field $\widetilde{\rho }/\rho _{0}$.
This requires solution of equation (\ref{wke}), \ie the calculation
of modulation. Since the precursor density fluctuations modulate a
state of interacting, finite amplitude Alfven waves, this constitutes
the unperturbed (unmodulated) state. Thus, the solution of equation
(\ref{wke}) is determined by a procedure similar to the Chapman-Enskog
expansion. To lowest order, then

\begin{equation}
\gamma _{k}^{\pm }N^{\pm }+C^{\pm }\left\{ N^{+},N^{-}\right\} \simeq 0\label{expans}\end{equation}
This relation implies that to lowest order, the linear growth rate
$\gamma ^{+}$ is in balance with the local nonlinear term $C^{+}$,
and the linear damping of counter streaming waves $\gamma ^{-}$ is
in balance with their generation by the scattering and conversion
of forward propagation waves. It is useful to note here that $C$
is in general a $2\times 2$ matrix operator, each component of which
is nonlinear in $N$. Formally, the solution of equation (\ref{expans})
defines the unmodulated wave streams driven by cyclotron resonance
of cosmic rays and inter-stream interaction, and damped by energy
coupling to small scales. To first order, then 

\begin{equation}
\frac{\partial }{\partial t}\widetilde{N}^{\pm }+\left(U+V_{A}\right)\frac{\partial }{\partial x}\widetilde{N}^{\pm }-kU_{x}\frac{\partial }{\partial k}\widetilde{N}^{\pm }-\left(\gamma ^{\pm }\widetilde{N}^{\pm }+\frac{\delta C}{\delta N}\widetilde{N}^{\pm }\right)=-\frac{kV_{A}}{2}\frac{\widetilde{\rho }_{x}}{\rho _{0}}\frac{\partial }{\partial k}\left\langle N^{\pm }\right\rangle \label{eq:Npmt}\end{equation}
gives the equation for the response of $\widetilde{N}$ to $\widetilde{\rho }_{x}/\rho $.
The problematical element of this equation is the last term of the
l.h.s. ($\gamma ^{\pm }\widetilde{N}^{\pm }+\left[\delta C/\delta N\right]\widetilde{N}^{\pm }$).
We write a 'Krook approximation' to this term as 

\begin{equation}
\gamma ^{\pm }\widetilde{N}^{\pm }+\frac{\delta C}{\delta N}\widetilde{N}^{\pm }\approx -\Delta \omega _{\mathbf{k}}\widetilde{N}^{\pm }\label{eq:Krook}\end{equation}
where $\Delta \omega _{\mathbf{k}}$ represents a nonlinear decay
or damping rate. We argue for the validity of this approximation by
noting that 

a.) since $\gamma ^{+}>0$, then $C<0$ is required for unperturbed
stationarity (\ie equation {[}\ref{expans}{]})

b.) since $C$ represents nonlinear interaction, $C$ must be nonlinear
in $N$, so $\widetilde{N}\delta C/\delta N>C$. For example, for
$C=-\alpha N^{2}$, $\gamma N-\alpha N^{2}=0$ gives the unperturbed
value $N_{0}=\gamma /\alpha $ and $\widetilde{N}\delta C/\delta N=-2\alpha N_{0}\widetilde{N}=-\gamma \widetilde{N}$

c.) so, $\gamma ^{\pm }\widetilde{N}^{\pm }+\left(\delta C/\delta N\right)\widetilde{N}^{\pm }<0$,
and hence corresponds to a \emph{damping} rate, which may be approximated
ala' Krook as $\Delta \omega _{k}\widetilde{N}^{\pm }$. Note that
this result is also consistent with the requirement of causality,
for $\Delta \omega _{k}\to \mathcal{O}\left(\epsilon \right)$. Note
here that the approximate form of the matrix $C$ is now diagonalized. 

It is tempting at this stage to take $\Delta \omega _{k}^{\pm }\simeq \left|\gamma ^{\pm }\right|$,
where the absolute value is required by consistency with causality.
However, this direct balance can be established \emph{only} for resonant
waves, whereas we are primarily interested in the larger wavelength
range $kr_{g}\left(p_{max}\right)<1$, where $\gamma \approx 0$.
Thus, $\Delta \omega _{k}$ is really due to nonlinear processes in
that part of the spectrum. However, to the extent that longer waves
are generated by the decay of shorter waves which are generated by
resonance, some link between $\Delta \omega _{k}$ and $\gamma _{k}$
persists. A derivation of $\Delta \omega _{k}$ which treats the effect
of strong nonlinear refraction which occurs on $kr_{g}<1$ scales
is presented later, in Sec.\ref{sec:Calculation-of-Wave}.

\begin{figure}
\epsscale{.25}
\plotone{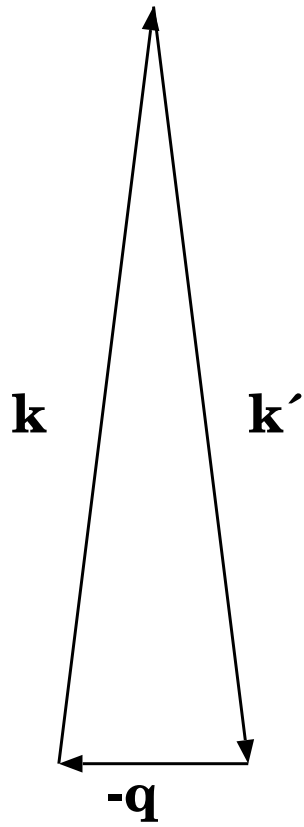}

\caption{Three-wave interaction of two nearly opposite Alfven quanta $\mathbf{k}$
and $\mathbf{k}^{\prime }$ with a long wave acoustic wave $\mathbf{q}$.\label{cap:Three-wave-interaction-of}}
\end{figure}

To calculate the refraction term we can now write equation (\ref{wke:av}),
linearized with respect to $\widetilde{N}^{\pm }$, as:

\begin{equation}
L^{\pm }\widetilde{N}^{\pm }=-kV_{A}\frac{\widetilde{\rho }_{x}}{2\rho _{0}}\frac{\partial }{\partial k}\left\langle N^{\pm }\right\rangle \label{L:oper}\end{equation}
 where

\[
L^{\pm }=\frac{\partial }{\partial t}+\left(U\pm V_{A}\right)\frac{\partial }{\partial x}-kU_{x}\frac{\partial }{\partial k}+\Delta \omega _{k}^{\pm }\]
is the linear propagator with an eddy damping rate $\Delta \omega _{k}^{\pm }$.
Solving equation (\ref{L:oper}) for $\widetilde{N}^{\pm }$, we thus
obtain the following mean field equation for $\left\langle N^{\pm }\right\rangle $

\begin{equation}
\frac{\partial }{\partial t}\left\langle N^{\pm }\right\rangle +U\frac{\partial }{\partial x}\left\langle N^{\pm }\right\rangle -kU_{x}\frac{\partial }{\partial k}\left\langle N^{\pm }\right\rangle -\frac{\partial }{\partial k}D_{k}\frac{\partial }{\partial k}\left\langle N^{\pm }\right\rangle =\gamma _{k}^{\pm }\left\langle N^{\pm }\right\rangle +\left\langle C\left(N^{\pm }\right)\right\rangle \label{wke:av2}\end{equation}
Notice that the mean field approximation to the refraction term in
the wave kinetic equation is a diffusion operator \emph{in} $k$ \emph{space}
for the Alfven wave spectrum. This diffusion represents \emph{random}
refraction by the acoustic perturbations $\widetilde{\rho }$ via
the density dependence of $V_{A}$ so

\begin{equation}
D_{k}=\frac{1}{4}k^{2}V_{A}^{2}\left\langle \frac{\widetilde{\rho }_{x}}{\rho _{0}}L^{-1}\frac{\widetilde{\rho }_{x}}{\rho _{0}}\right\rangle \label{diff:op}\end{equation}
 $D_{k}$ is an example of the well-known phenomenon of \emph{induced
diffusion}. Induced diffusion is a generic type of 3-wave interaction
due to resonant triads with $\mathbf{k+k}^{\prime }\mathbf{+q}=0$
but $\left|\mathbf{k}\right|,\left|\mathbf{k}^{\prime }\right|\gg \left|\mathbf{q}\right|$.
Such triads can be represented by thin, isosceles triangles, Fig.\ref{cap:Three-wave-interaction-of}.
The basic physics of induced diffusion is random refraction by large
scale perturbations. Induced diffusion can thus be obtained from eikonal
theory approaches. The inversibility which ultimately underpins the
diffusion has its origins in the stochasticity of the Alfven wave
rays in the field of density perturbations.  Transforming to Fourier
space, we first represent $\widetilde{\rho }$ as

\[
\widetilde{\rho }=\sum _{q}\widetilde{\rho }_{q}e^{iqx-i\Omega _{q}t}\]
 and note that due to the local Galilean invariance of $L$, we can
calculate its Fourier representation in the reference frame moving
with the plasma at the speed $U(x)$ as:

\begin{equation}
L_{k,q}^{\pm }=\pm iqV_{A}+\Delta \omega _{k}^{\pm }-kU_{x}\frac{\partial }{\partial k}\label{L:op:four}\end{equation}
 Then, equation (\ref{diff:op}) can be re-written as: 

\begin{equation}
D_{k}=\frac{1}{2}k^{2}V_{A}^{2}\sum _{q}q^{2}\left|\frac{\widetilde{\rho }_{q}}{\rho _{0}}\right|^{2}\Re L_{k,q}^{-1}\label{diff:op2}\end{equation}
 The last (wave refraction) term on the r.h.s. of equation (\ref{L:op:four})
can be estimated as $U_{1}^{2}/\kappa (p_{max})$, which is the inverse
acceleration time and can be neglected in comparison to the frequencies
$qV_{A}$ and $\Delta \omega $. Hence, for $\Re L_{k,q}^{\pm -1}$
we have:

\begin{equation}
\Re L_{k,q}^{\pm -1}\approx \frac{\Delta \omega _{k}^{\pm }}{q^{2}V_{A}^{2}+\Delta \omega _{k}^{\pm 2}}\label{eq:ReL}\end{equation}
Note that equation (\ref{eq:ReL}) states that the correlation time
$\tau _{c}$ for induced diffusion is set by the Alfven wave damping
time, ($\sim 1/\Delta \omega _{k}$) reduced by the effects of finite
transit time for a wave to propagate through the density perturbation,
if $\tau _{c}qV_{A}>1$. Clearly, equation (\ref{eq:ReL}) has two
limits, a 'strong turbulence' limit where $\Re L_{k,q}^{\pm -1}\simeq \Delta \omega _{k}^{-1}=\tau _{ck}$
and a 'weak turbulence limit' where $\Re L_{k,q}^{\pm -1}\simeq 1/q^{2}V_{A}^{2}\tau _{ck}$.
For further convenience, we write the population density of acoustic
waves (phonons) as 

\[
N_{q}^{s}=\frac{W_{q}}{\omega _{q}^{s}}\]
 where $W_{q}$ is the energy density of acoustic waves (with $\omega _{q}^{s}=qC_{s}$
their frequency).

\[
W_{q}=C_{s}^{2}\frac{\widetilde{\rho }_{q}^{2}}{\rho _{0}}\]
 For $D_{k}$ in equation (\ref{wke:av2}) we thus finally have

\begin{equation}
D_{k}=\frac{k^{2}V_{A}^{2}}{4C_{s}^{2}\rho _{0}}\sum _{q}q^{2}\omega _{q}^{s}\frac{\Delta \omega _{k}^{\pm }}{q^{2}V_{A}^{2}+\Delta \omega _{k}^{\pm 2}}N_{q}^{s}\label{Dk1}\end{equation}
 Note that $D_{k}$ represents the rate at which the wave vector of
the Alfven wave random walks due to stochastic refraction. Of course,
such a random walk necessarily must generate larger scales (smaller
$k$), thus in turn allowing the confinement of higher energy particles
to the shock. Thus, confinement of higher energy particles is a natural
consequence of random Alfven wave refraction in acoustic perturbations.

\subsection{Acoustic turbulence}

In contrast to the Alfvenic turbulence that originates in the shock
precursor due to cyclotron emission from accelerated particles, there
are (at least) two separate sources of long wave acoustic perturbations.
One is related to parametric and modulational \markcite{SagdGal69,Skill75b}({Sagdeev} \& {Galeev} 1969; {Skilling} 1975)
processes undergone by the Alfven waves. These take the usual form
of decay of an Alfven wave into another Alfven wave and an acoustic
wave. The other source is the pressure gradient of CRs, which \emph{directly}
drives \emph{linear instability}. The latter leads to emission of
unstable sound waves via the Drury instability. By analogy with equation
(\ref{wke:av}), we can then write the following wave kinetic equation
for the acoustic waves:

\[
\frac{\partial }{\partial t}N_{q}+U\frac{\partial }{\partial x}N_{q}-qU_{x}\frac{\partial }{\partial k}N_{q}=\left(\gamma _{q}^{d}+\gamma _{D}\right)N_{q}+C\left\{ N_{q}\right\} \]
 Here $\gamma _{D}$ is the Drury instability growth rate $\gamma _{D}=\gamma _{D}\left[\nabla P_{CR}\right]$
and $\gamma _{q}^{d}$ is the growth rate of the decay instability
$\gamma _{q}^{d}=\gamma _{q}^{d}\left[N_{k}\right]$. We first consider
the decay instability of Alfven waves. Note, however, that the combination
of Drury instability and decay instability can lead to more rapid
generation of mesoscale fields by coupling together the linear and
nonlinear processes, both of which amplify density perturbations.

\subsubsection{Modulational Instability of Alfven Wave Packets in a Density Scatterer
Field}

The mechanism of this instability is the growth of the density (acoustic)
perturbations due to the action of the ponderomotive force on the
acoustic waves by the Alfven waves. This force can be regarded as
an effective radiation pressure term which must appear in the hydrodynamic
equation of motion for the sound waves (written below in the comoving
plasma frame)

\[
\frac{\partial \widetilde{V}}{\partial t}=-\frac{1}{\rho _{0}}\frac{\partial }{\partial x}\left(c_{s}^{2}\widetilde{\rho }+\widetilde{P}_{rad}\right)\]
 Eliminating velocity by making use of continuity equation

\[
\frac{\partial \widetilde{\rho }}{\partial t}+\rho _{0}\frac{\partial \widetilde{V}}{\partial x}=0,\]
 we obtain

\begin{equation}
\frac{\partial ^{2}\widetilde{\rho }}{\partial t^{2}}=\frac{\partial ^{2}}{\partial x^{2}}\left(c_{s}^{2}\widetilde{\rho }+\widetilde{P}_{rad}\right)\label{sound:eq}\end{equation}
 The Alfven wave pressure can be related to their population densities
via the energy density, so 

\[
P_{rad}=\sum _{k}\omega _{k}\left(\widetilde{N}^{+}+\widetilde{N}^{-}\right)\]
 Using the relation (\ref{L:oper}) between the density perturbations
and the Alfven waves and separating the populations of forward and
backward propagating sound waves $\rho ^{\pm }$, we can obtain from
equation (\ref{sound:eq}) the following dispersion relation for the
nonlinear growth of the the acoustic branch via modulational interactions

\[
\omega ^{2}-q^{2}C_{s}^{2}=q^{2}\sum _{k}\frac{\omega _{k}}{2\rho _{0}}iqkV_{A}L_{k,q}^{\pm -1}\frac{\partial }{\partial k}\left\langle N_{k}^{\pm }\right\rangle \]
 upon writing $\omega =\pm qc_{s}+i\gamma ^{\pm }$, we then find
the following growth rate of acoustic perturbations

\[
\gamma ^{\pm }=\frac{q^{2}}{4\rho _{0}}\frac{V_{A}}{c_{s}}\sum _{k}k\omega _{k}L_{k,q}^{\pm -1}\frac{\partial }{\partial k}\left\langle N_{k}^{\pm }\right\rangle \]
 Note that modulational instability requires an inverted population
of Alfven quanta ie $\partial \left\langle N\right\rangle /\partial k>0$.
As Alfven waves are generated by high energy resonant particles in
a limited band of $k$ space at short wavelength (\ie $kr_{g}\sim 1$),
such an inversion clearly can occur. Note also that shorter wavelength
modulations appear to grow faster, but this trend saturates when $q\ga \Delta \omega _{k}/V_{A}$.
Note also that the coherence time between the Alfvenic packet and
the modulating density perturbation field is set by $L_{k,q}^{-1}\sim \Delta \omega _{k}/\left(q^{2}V_{gr}^{2}+\Delta \omega _{k}^{2}\right)$
which, not surprisingly, also sets the correlation time in $D_{k}$.

\subsubsection{Drury Instability}

There is a linear instability which also leads to efficient generation
of acoustic waves and is driven by the pressure gradient of the CRs
in the shock precursor. Such a $\nabla P_{CR}$-driven process is
of particular interest, as it taps free energy which is generated
and stored as a consequence of the acceleration process, itself. The
growth rate has been calculated by \markcite{DruryCOSP84}{Drury} (1984) and \markcite{DruryFal86}{Drury} \& {Falle} (1986)
(see also \markcite{ZankAM90,KangJR92}{Zank}, {Axford}, \& {McKenzie}, 1990; {Kang}, {Jones}, \& {Ryu}, 1992), and can be written in the
form:

\begin{equation}
\gamma _{D}^{\pm }=-\frac{\gamma _{C}P_{C}}{\rho \kappa }\pm \frac{P_{Cx}}{C_{s}\rho }\left(1+\frac{\partial \ln \kappa }{\partial \ln \rho }\right)\label{DrGrRate}\end{equation}
 Here $P_{C}$ and $P_{Cx}$ are the CR pressure and its derivative,
respectively, and $\gamma _{C}$ is their adiabatic index. Note that
instability requires $d\ln \kappa /d\ln \rho >-1$, so that the structure
and dependencies of the cosmic ray diffusivity are critical to the
Drury instability. The physics of this instability is that the density
fluctuations in the decelerating precursor flow induce the CR pressure
gradient fluctuations which (unless they are \emph{exactly} proportional
to the density fluctuations) make the flow deceleration non-uniform
and so can amplify the initial density fluctuations (~\ie\markcite{DruryCOSP84,KangJR92}{Drury}, 1984; {Kang} {et~al.}, 1992).
For an efficiently accelerating shock, the adiabatic index $\gamma _{C}$
in equation (\ref{DrGrRate}) is $\gamma _{C}\approx 4/3$. Note that
we have omitted a term $-U_{x}$ which is related to simple compression
of wave number density by the flow, and which would enter the r.h.s.
of equation (\ref{sound:eq}) (see \markcite{DruryFal86}{Drury} \& {Falle}, 1986). This term
is smaller, by a factor of $C_{s}/U$, than the second (destabilizing)
term. The first term is damping caused by CR diffusion, calculated
earlier by \markcite{Ptuskin81}{Ptuskin} (1981).

\subsection{Nonlinear Wave Trains and Alfven--Acoustic Coupling}

Compressibility of the media in a shock environment is believed to
be responsible for the formation of coherent magnetic structures observed
upstream of the Earth's bow shock and interplanetary shocks. These
structures are thought to evolve from phase steepened Alfven wave
trains and should be relevant to the SNR shocks as well. The bottom
line of this highly evolved and mature field of research is that parallel
compressibility transforms Alfven wave trains into steepened Alfven
solitons, described by the DNLS (Derivative Nonlinear Schr\"{o}dinger
Equation) \markcite{Kennel88,Medved99}({Kennel} {et~al.} 1988a; {Medvedev} 1999)and its variants. The DNLS describes,
within the framework of coherent interactions, the refraction of Alfven
waves by density perturbations and its feedback on the Alfven wave
envelope. Thus, the DNLS describes a process which is the coherent
analogue of $k$- space diffusion and which does not suffer from the
limitations of eikonal theory. A simplified derivation of the DNLS
is presented below. Consider the familiar Alfven wave dispersion law
$\omega =k_{\parallel }V_{A}$. It stems from the differential equation
for the magnetic field perturbations. If we now entertain the possibility
of slowly varying density perturbations, we have

\begin{equation}
\frac{\partial \widetilde{\mathbf{B}}}{\partial t}=\frac{1}{2}\frac{\partial }{\partial z}V_{A}\frac{\widetilde{\rho }}{\rho _{0}}\widetilde{\mathbf{B}}\label{AlfWaveDisp}\end{equation}
 where $\widetilde{\mathbf{B}}$ is the magnetic field perturbation
envelope function in the Alfven wave, and $\widetilde{\rho }$ is
the density perturbation induced by parallel compression. From the
continuity and parallel momentum equations we have

\begin{eqnarray}
\frac{\partial \widetilde{\rho }}{\partial t} & = & -\rho _{0}\frac{\partial \widetilde{V}_{z}}{\partial z}\label{eq:rhotilde}\\
\rho _{0}\frac{\partial \widetilde{V}_{z}}{\partial z} & = & -\frac{\partial }{\partial z}\left(\widetilde{p}+\frac{\widetilde{B}^{2}}{8\pi }+\frac{\rho \widetilde{V}^{2}}{2}\right)\label{eq:Vztilde}
\end{eqnarray}
with $\rho \widetilde{V}^{2}/2\sim \widetilde{B}^{2}/8\pi $ (\ie
both $\widetilde{V}$ and $\widetilde{B}$ enter ala' Bernoulli) and
$\widetilde{V}\sim \widetilde{B}$ since the basic waves are Alfven
waves. Assuming that the perturbations propagate at approximately
the Alfven velocity, from equations (\ref{eq:rhotilde}-\ref{eq:Vztilde})
we obtain the relation between wave envelope perturbation levels and
density fluctuation levels, \ie

\[
\frac{\widetilde{\rho }}{\rho _{0}}\simeq -\frac{1}{2}\left(1-C_{s}^{2}/V_{A}^{2}\right)^{-1}\frac{\widetilde{B}^{2}}{B_{0}^{2}}\]
Note that this relation gives a clear and simple relation between
density and magnetic fluctuation energy levels. Substituting this
in equation (\ref{AlfWaveDisp}) we obtain

\[
\frac{\partial \widetilde{\mathbf{B}}}{\partial t}+\frac{1}{4}\frac{\partial }{\partial z}\left(\frac{V_{A}}{1-C_{s}^{2}/V_{A}^{2}}\right)\frac{\left|\widetilde{\mathbf{B}}\right|^{2}}{B_{0}^{2}}\widetilde{\mathbf{B}}=0\]
 Dispersive corrections to the Alfven mode can be easily added into
this equation as well. These ultimately limit steepening, producing
collisionless shocks. For the singular case $\beta \approx 1$, a
modified kinetic equation (KNLS) can be derived \markcite{MedvDiam96}({Medvedev} \& {Diamond} 1996).
The DNLS equation possesses soliton solutions that may self-organize
in quasi periodic structures (wavetrains). Upon addition of dissipative
term and a driver to this equation (\eg  growth due to instability),
quasi periodic shock train solutions may also be obtained\markcite{KennelJetp88,Hada90}({Kennel} {et~al.} 1988b; {Hada} {et~al.} 1990).
Such DNLS and KNLS shocks can trap and mirror energetic particles,
thus enhancing their confinement. Consideration of DNLS/KNLS structures
and their dynamics gives a clear and compelling physical picture of
the Alfvenic mirroring process as well as constituting a coherent
analogue of the stochastic modulation discussed above. In particular,
the coherent modulation of instability amplifies the magnetic energy
of the \emph{envelope}, and thus couples energy to large scales.

\section{Wave Quanta Evolution in Strongly Refracting Turbulence\label{sec:Calculation-of-Wave}}

Noting that Alfven waves are generated at $kr_{g}\sim 1$, while the
modulational interaction which scatters the spectral population to
large scales occurs at $kr_{g}<1$. In the likely case that this interaction
is strong (\ie $\widetilde{\rho }/\rho _{0}\sim 1$) and stochastic,
it can directly produce wave packet decorrelation, which also contributes
to the interaction damping rate $\Delta \omega _{k}$. Here, we calculate
this decorrelation. Since in this case, stochastic refraction is dominant,
the details of the l.h.s. of the wave kinetic equation are not important,
so we write, in the precursor flow frame

\begin{equation}
\frac{\partial N^{\pm }}{\partial t}+\left(\left\langle u\right\rangle \pm V_{A}\right)\frac{\partial N^{\pm }}{\partial x}-k\frac{\partial \left\langle u\right\rangle }{\partial x}\frac{\partial N^{\pm }}{\partial k}+\frac{\partial }{\partial x}\widetilde{u}\widetilde{N}^{\pm }+\frac{\partial }{\partial k}kV_{A}\frac{\widetilde{\rho }_{x}}{\rho _{0}}\widetilde{N}^{\pm }=0\label{app1:wke}\end{equation}
The mean field equation is just 

\begin{equation}
\frac{\partial \left\langle N^{\pm }\right\rangle }{\partial t}+\left(\left\langle u\right\rangle \pm V_{A}\right)\frac{\partial \left\langle N^{\pm }\right\rangle }{\partial x}-k\frac{\partial \left\langle u\right\rangle }{\partial x}\frac{\partial \left\langle N^{\pm }\right\rangle }{\partial k}+\frac{\partial }{\partial x}\left\langle \widetilde{u}\widetilde{N}^{\pm }\right\rangle +\frac{\partial }{\partial k}kV_{A}\left\langle \frac{\widetilde{\rho }_{x}}{\rho _{0}}\widetilde{N}^{\pm }\right\rangle =0\label{app:wke1}\end{equation}
Note that the problem of determining the evolution of the mean wave
populations reduces to calculating the correlations $\left\langle \widetilde{u}\widetilde{N}^{\pm }\right\rangle $
and $\left\langle \left(\widetilde{\rho }_{x}/\rho _{0}\right)\widetilde{N}^{\pm }\right\rangle $,
which constitute the spatial and wavenumber fluxes, respectively.
In the spirit of quasilinear theory, we calculate the value of $\widetilde{N}^{\pm }$
to close the correlators by determining the response of $N^{\pm }$
to the acoustic wave perturbations $\widetilde{u}$ and $\widetilde{\rho }_{x}/\rho _{0}$.
Writing the linear propagator

\begin{equation}
\left(\left\langle u\right\rangle \pm V_{A}\right)\frac{\partial }{\partial x}-k\frac{\partial \left\langle u\right\rangle }{\partial x} \frac{\partial }{\partial k} \equiv L_{\pm },\label{app1:Lpm}\end{equation}
the response is given by

\[
\frac{\partial \widetilde{N}^{\pm }}{\partial t}+L_{\pm }\widetilde{N}^{\pm }+C\left[\widetilde{N}^{\pm }\right]=-\widetilde{u}\frac{\partial \left\langle N^{\pm }\right\rangle }{\partial x}-\frac{kV_{A}}{2}\frac{\widetilde{\rho }_{x}}{\rho _{0}}\frac{\partial }{\partial k}\left\langle N^{\pm }\right\rangle \]
Note that the Hamiltonian structure of eikonal theory means that the
flow in $(\mathbf{x},\mathbf{k})$ space is incompressible, thus allowing
re-arrangement of the order of the derivatives. Here $C\left[\widetilde{N}^{\pm }\right]$
is that portion of the nonlinear terms which is phase coherent with
the population fluctuation. Taking modulations to have the form

\[
\left(\begin{array}{c}
 \widetilde{u}\\
 \widetilde{\rho }/\rho _{\pm }\\
 \widetilde{N}^{\pm }\end{array}
\right)=\sum _{q}\left(\begin{array}{c}
 \widetilde{u}_{q}\\
 \widetilde{\rho }_{q}/\rho _{0}\\
 \widetilde{N}_{q}^{\pm }\end{array}
\right)e^{i\left(qx-\Omega t\right)}\]
then the response equation becomes

\[
-i\left(\Omega -qL_{q}\right)\widetilde{N}_{q,\Omega }^{\pm }+C\left[\widetilde{N}^{\pm }\right]_{q,\Omega }\equiv -\widetilde{u}_{q}\frac{\partial \left\langle N^{\pm }\right\rangle }{\partial x}-\frac{kV_{A}}{2}iq\frac{\widetilde{\rho }_{q}}{\rho _{0}}\frac{\partial }{\partial k}\left\langle N^{\pm }\right\rangle \]
Here we have re-written $L$ as $iqL_{q}$ for the corresponding Fourier
mode. $C\left[\widetilde{N}^{\pm }\right]_{q,\Omega }$ is given explicitely
by (we introduce a {}``two-vector'' $\bar{q}\equiv \left(q,\Omega \right)$
for short)

\begin{eqnarray*}
C\left[\widetilde{N}^{\pm }\right]_{\bar{q}} & = & -\left[\frac{\partial }{\partial x}\left(\widetilde{u}\widetilde{N}^{\pm }\right)+\frac{\partial }{\partial k}kV_{A}\left(\frac{\widetilde{\rho }_{x}}{\rho _{0}}\widetilde{N}^{\pm }\right)\right]_{\bar{q}}\\
 & = & \sum _{\bar{q}^{\prime }}\left[-iq\widetilde{u}_{-\bar{q}^{\prime }}\widetilde{N}_{\bar{q}^{\prime }+\bar{q}}^{\pm }+\frac{\partial }{\partial k}\frac{kV_{A}}{2}iq^{\prime }\frac{\widetilde{\rho }_{-\bar{q}^{\prime }}}{\rho _{0}}\widetilde{N}_{\bar{q}^{\prime }+\bar{q}}^{\pm }\right]\\
 & = & \Delta \omega _{\bar{q}}\widetilde{N}_{\bar{q}}^{\pm }
\end{eqnarray*}
The expression for $C\left[\widetilde{N}^{\pm }\right]_{q,\Omega }$
may, in turn, be obtained as in quasilinear theory by approximating
$\widetilde{N}_{\bar{q}^{\prime }+\bar{q}}^{\pm }$ in terms of the
beats at $\bar{q}+\bar{q}^{\prime }$ determined by the wave kinetic
equation, \ie

\[
-i\left[\Omega +\Omega ^{\prime }-\left(qL_{q}+q^{\prime }L_{q^{\prime }}\right)\right]\widetilde{N}_{\bar{q}+\bar{q}^{\prime }}^{\pm }+\Delta \omega _{\bar{q}+\bar{q}^{\prime }}\widetilde{N}_{\bar{q}+\bar{q}^{\prime }}^{\pm }=iq\widetilde{u}_{\bar{q}^{\prime }}\widetilde{N}_{\bar{q}}^{\pm }+\frac{kV_{A}}{2}iq^{\prime }\frac{\widetilde{\rho }_{\bar{q}^{\prime }}}{\rho _{0}}\frac{\partial }{\partial k}\widetilde{N}_{\bar{q}}^{\pm }\]
A formal solution then yields

\[
\widetilde{N}_{\bar{q}+\bar{q}^{\prime }}^{\pm }=R_{\bar{q}+\bar{q}^{\prime }}\left(iq\widetilde{u}_{\bar{q}^{\prime }}\widetilde{N}_{\bar{q}}^{\pm }+\frac{kV_{A}}{2}iq^{\prime }\frac{\widetilde{\rho }_{\bar{q}^{\prime }}}{\rho _{0}}\frac{\partial }{\partial k}\widetilde{N}_{\bar{q}}^{\pm }\right)\]
where

\[
R_{\bar{q}+\bar{q}^{\prime }}^{-1}=-i\left[\Omega +\Omega ^{\prime }-\left(qL_{q}+q^{\prime }L_{q^{\prime }}\right)\right]+\Delta \omega _{\bar{q}+\bar{q}^{\prime }}\]
This in turn gives the result

\begin{eqnarray*}
C\left[\widetilde{N}^{\pm }\right]_{\bar{q}} & = & \Delta \omega _{\bar{q}}\widetilde{N}_{\bar{q}}^{\pm }\\
 & = & \sum _{\bar{q}^{\prime }}\left(-iq\widetilde{u}_{-\bar{q}^{\prime }}-\frac{\partial }{\partial k}\frac{kV_{A}}{2}iq^{\prime }\rho _{-\bar{q}^{\prime }}\right)R_{\bar{q}+\bar{q}^{\prime }}\\
 & + & \left(iq\widetilde{u}_{\bar{q}^{\prime }}\widetilde{N}_{\bar{q}}^{\pm }+\frac{kV_{A}}{2}iq^{\prime }\frac{\widetilde{\rho }_{\bar{q}^{\prime }}}{\rho _{0}}\frac{\partial }{\partial k}\widetilde{N}_{\bar{q}}^{\pm }\right)
\end{eqnarray*}
Ignoring cross terms the expression for $\Delta \omega _{\bar{q}}\widetilde{N}_{\bar{q}}^{\pm }$
may be simplified to 

\[
\Delta \omega _{\bar{q}}\widetilde{N}_{\bar{q}}^{\pm }=q^{2}D_{\bar{q}}\widetilde{N}_{\bar{q}}-\frac{\partial }{\partial k}D_{\bar{q},k}\frac{\partial \widetilde{N}_{\bar{q}}}{\partial k}\]
where

\begin{eqnarray*}
D_{\bar{q}} & = & \sum _{\bar{q}^{\prime }}\left|\widetilde{u}_{\bar{q}^{\prime }}\right|^{2}\Re R_{\bar{q}^{\prime }+\bar{q}}\\
D_{\bar{q},k} & = & \sum _{\bar{q}^{\prime }}\frac{k^{2}V_{A}^{2}}{4}q^{\prime 2}\left(\frac{\widetilde{\rho }_{\bar{q}^{\prime }}}{\rho _{0}}\right)^{2}\Re R_{\bar{q}^{\prime }+\bar{q}}
\end{eqnarray*}

correspond to wave packet diffusion in position and $k$, respectively.
Note also that 

\[
\Re R_{\bar{q}^{\prime }+\bar{q}}=\frac{\Delta \omega _{\bar{q}+\bar{q}^{\prime }}}{\left[\Omega +\Omega ^{\prime }-\left(qL_{q}+q^{\prime }L_{q^{\prime }}\right)\right]^{2}+\Delta \omega _{\bar{q}+\bar{q}^{\prime }}^{2}}\]
so that $\Delta \omega _{q}$ is defined recursively, as usual, and
that both types of scattering contribute to the total decorrelation.
Finally, again ignoring cross terms, we can now obtain the mean field
equation, which is 

\begin{eqnarray*}
\frac{\partial \left\langle N^{\pm }\right\rangle }{\partial t}+\left(\left\langle u\right\rangle \pm V_{A}\right)\frac{\partial \left\langle N^{\pm }\right\rangle }{\partial x}-k\frac{\partial \left\langle u\right\rangle }{\partial x}\frac{\partial \left\langle N^{\pm }\right\rangle }{\partial k} &  & \\
-\frac{\partial }{\partial x}D\frac{\partial }{\partial x}\left\langle N^{\pm }\right\rangle -\frac{\partial }{\partial k}D_{k}\frac{\partial }{\partial k}\left\langle N^{\pm }\right\rangle  & =0 & 
\end{eqnarray*}

Here $D$ and $D_{k}$ again correspond to spatial and wavenumber
diffusion of an Alfven wave packet, and are given by $\lim _{\bar{q}\to 0}D_{\bar{q}},D_{k,\bar{q}}$,
respectively. The output of this relatively straightforward calculation
is the set of renormalized quasilinear diffusion coefficients $D$,
$D_{k}$. In each of these the decorrelation rate is set by decorrelation
rate $\Delta \omega _{q}$, since the dissipation in the acoustic
wave spectrum is weak. Since $\Delta \omega $ is defined recursively,
explicit relations between $\Delta \omega _{k}$ and the refracting
field spectrum are best discussed in various simple limits. In the
case that decorrelation is dominated by spacial scattering, 

\[
\Delta \omega _{q}=q^{2}\sum _{\bar{q}^{\prime }}\left|\widetilde{u}_{\bar{q}^{\prime }}\right|^{2}\frac{\Delta \omega _{\bar{q}+\bar{q}^{\prime }}}{\bar{\Omega }^{2}+\Delta \omega _{\bar{q}+\bar{q}^{\prime }}^{2}}\]
Here $\bar{\Omega }$ absorbs the propagator Doppler shift. In this
case then, $\Delta \omega \sim \left(q^{2}\left\langle \widetilde{u}^{2}\right\rangle -\bar{\Omega }^{2}\right)^{1/2}$.
Note that a critical level of $\left\langle \widetilde{u}^{2}\right\rangle $
is required for irreversibility, \ie to make a stochastic Doppler
shift sufficient to overcome the frequency $\bar{\Omega }$. In the
limit where wave-number scattering dominates but the waves are weakly
dispersive (\ie $dV_{g}/dk\to 0$)

\[
\Delta \omega _{q}\simeq \frac{1}{\left(\Delta k\right)^{2}}\sum _{\bar{q}^{\prime }}\frac{k^{2}V_{A}^{2}}{4}\bar{q}^{2}\left|\frac{\widetilde{\rho }_{q}}{\rho _{0}}\right|^{2}\frac{\Delta \omega _{q}}{\bar{\Omega }^{2}+\Delta \omega _{q}^{2}}\]
so

\[
\Delta \omega \sim \left[\frac{1}{\left(\Delta k\right)^{2}}\frac{k^{2}V_{A}^{2}}{4}\frac{\left\langle \left(\nabla \widetilde{\rho }\right)^{2}\right\rangle }{\rho _{0}^{2}}-\bar{\Omega }^{2}\right]^{1/2}.\]
Again, a critical rms density gradient fluctuation level is required.
In the limit of stronger dispersion and strong turbulence, the decorrelation
rate is 

\[
\Delta \omega _{q}\sim \left(q^{2}V_{g}^{\prime 2}\frac{k^{2}V_{A}^{2}}{4}\frac{\left\langle \left(\nabla \widetilde{\rho }\right)^{2}\right\rangle }{\rho _{0}^{2}}\right)^{1/4}\]
Note that in this case $\Delta \omega \sim \left(\nabla \rho /\rho _{0}\right)_{rms}^{1/2}$.
It is important to stress that here, \emph{the physical processes
represented by} $\Delta \omega _{q}$ \emph{are random advection and
refraction of wave packets by the acoustic wave fluctuations on scales}
$q$, \emph{where} $qr_{g}\ll 1$, and \emph{not} the interaction
of Alfven waves with $kr_{g}\sim 1$. Since, of course, \emph{both}
processes occur, and since the modulational interaction of Alfven
wave packets with density perturbations generates larger scale waves,
the physics of wave packet decorrelation is, not surprisingly, \emph{strongly}
dependent on scale. While this may result in some technical difficulties
in qualitative calculations, it \emph{does} ensure that irreversible
'modulational turbulence' dynamics persists over a broad range of
scales, and is \emph{not sharply localized at} $kr_{g}\sim 1$.

\section{Mechanisms of Magnetic Energy Transfer to Larger Scales\label{sec:Mechanisms-of-Magnetic}}

As should be clear from the considerations discussed above, there
are a variety of nonlinear processes that can lead to the transfer
of magnetic energy (generated by accelerated particles in form of
the resonant Alfven waves) to longer scales. First, as it can be seen
from equation (\ref{wke:av2}) (\ie last term on the l.h.s.), scattering
of the Alfven waves in $k$ space due to acoustic perturbations transfers
magnetic fluctuation energy away from the resonant excitation region
to smaller (and also to larger) $k$, and also amplifies the long
wavelength acoustic scattering field. Second, the nonlinear interaction
of Alfven waves and magnetosonic waves represented by the wave collision
term on the r.h.s. can drive such a process. Next, as we discussed
in the last subsection, solutions of DNLS-KNLS equations can be well
represented by quasiperiodic wave structures. The interaction between
such structures, such as, for example, shock waves in the Burgers
model, leads to their coalescence, which in turn, means an efficient
transfer of excitation to larger scales. Finally, we remind the reader
that even within the frame work of weak turbulence theory, induced
scattering (\ie  nonlinear Landau damping) of Alfven waves on thermal
protons leads to a systematic decrease in the energy of quanta which,
given the dispersion law, again means energy scattering to longer
wavelength.

Continuing our main discussion of the wave refraction by acoustic
perturbations generated by the Drury and modulational instabilities,
it is important to emphasize the following point. As seen from the
instability growth rate, equation (\ref{DrGrRate}), the Drury instability
growth rate is proportional to the gradient of $P_{c}$. While one
would naively expect Drury instability to \emph{relax} the gradient
that drives it, we note that the outcome of the dynamic feedback we
discuss here suggests that some of the relaxation would be off-set
enhanced confinement and acceleration. Indeed, this might ultimately
reinforce the instability, possibly triggering bursts or cyclic growth.
This might help realize mechanisms of regulation of $P_{c}$ suggested
earlier in \markcite{mdv00}({Malkov} {et~al.} 2000). 

We proceed, however, with a simpler approach by not treating the acoustic
perturbations fully self-consistently \ie  we do not consider the
connection between acoustics and particle acceleration and shock modification.
Instead, at this level we treat them as a developed to some quasi-stationary
state as a result of \eg Drury instability and we consider how the
Alfvenic turbulence then evolves in their field. Thus, we address
only the simplest and most fundamental problem, that of the evolution
of the Alfven wave population in a field of ambient density perturbations.

\subsection{Estimates of Alfven Wave Diffusion in $k$ space\label{sub:Estimates-of-Alfven}}

The spectral evolution of the Alfven waves generated by accelerated
particles is crucial for confinement and further acceleration of these
particles. It is described by equation (\ref{wke:av2}) and involves
several processes: 

\begin{enumerate}
\item wave generation at the rate $\gamma $ 
\item convection of the waves to the shock from the upstream side at the
flow speed $U$
\item blue-shift of the waves to short scales by the flow compression (term
$\propto U_{x}$, see \markcite{mdj02}{Malkov} {et~al.}, 2002) 
\item nonlinear interaction (term $\propto C\left(N\right)$) with each
other and particles (induced scattering)
\item random scattering in wave number on acoustic perturbations (diffusion
in $k$) 
\end{enumerate}
Clearly, an exact solution of the problem is very difficult, particularly
because all these phenomena are coupled and related and, ideally,
should be treated self-consistently. For example, the wave turbulence
level $N\left(k\right)$ directly enters $\kappa \left(p\right)$
of the convection-diffusion equation (\ref{dc1}), the solution of
which determines both the flow profile $U\left(x\right)$ (through
$P_{c}$ and eqs.{[}\ref{mas:c}-\ref{mom:c}{]}), and the growth
rate $\gamma $ (through $P_{c}$) and also the Alfven quanta scattering
rate $D_{k}$ (through the Drury or parametric instabilities). Each
of these appears explicitly in the wave kinetic equation, thus closing
the feedback loop. A fully self-consistent level of description would
jump several steps ahead of the the current DSA models and is clearly
beyond the scope of this paper. Therefore, to achieve any simple understanding,
several simplifications are necessary. In particular, we wish to focus
here on the spreading of the wave population in $k$ space, assuming
that the spectrum is initially generated by accelerated particles
upstream of the shock in the resonant $k$ domain. Therefore, we ignore
nonlinear interaction between waves, as well as wave refraction described
by the flow compression $U_{x}$. This effect, and (partially) also
the wave self-nonlinearity have been considered by \markcite{mdj02}{Malkov} {et~al.} (2002). 

First, we rewrite equation (\ref{wke:av2}) in the following simplified
form

\begin{equation}
\frac{\partial I_{k}}{\partial t}+U\frac{\partial I_{k}}{\partial x}-\frac{\partial }{\partial k}D_{k}\frac{\partial I_{k}}{\partial k}=\frac{2U}{kV_{A}}\frac{\partial P}{\partial x}\label{wkeI}\end{equation}
 where we have introduced the dimensionless wave intensity $I_{k}$
normalized to the background magnetic energy

\[
I_{k}=\frac{\left|B_{k}\right|^{2}}{B_{0}^{2}}\]
along with the partial particle pressure $P\left(x,p\right)$ in the
$p\gg 1$ range (see equation {[}\ref{Pc}{]} ) normalized to the
upstream ram pressure $\rho _{1}u_{1}^{2}$

\begin{equation}
P=\frac{4\pi }{3}\frac{mc^{2}}{\rho _{1}u_{1}^{2}}p^{4}f\left(p,x\right)\label{Ppart}\end{equation}
We do not distinguish here between the forward and backward backward
propagating waves $N^{\pm }$ or Alfven and magnetosonic waves. Rather,
we simply incorporate \emph{all} types of magnetic fluctuations in
the spectral density $I_{k}$. The wave induced diffusivity $D_{k}$
from equation (\ref{Dk1}) can be represented as 

\begin{equation}
D_{k}=\frac{1}{4}k^{2}V_{A}^{2}\sum _{q}\frac{q^{2}\Delta \omega }{q^{2}V_{g}^{2}+\Delta \omega ^{2}}\frac{\widetilde{\rho }_{q}^{2}}{\rho _{0}^{2}}\label{Dk2}\end{equation}
To further simplify this expression we note that there are several
reasons for the resonance broadening $\Delta \omega $ in the last
expression, as discussed earlier. One is the nonstationarity of the
Alfven waves being scattered by the acoustic fluctuations, which yields$\Delta \omega \sim \gamma $.
A different reason is the refractive scattering itself. It comes in
two flavors, as discussed earlier, one of which is related to the
dispersion of the group velocity, $V_{g}^{\prime }=\partial V_{g}/\partial k$,
acting in concert with the scattering in $k$, namely $\Delta \omega \sim \left(qV_{g}^{\prime }\right)^{2/3}D_{k}^{1/3}$,
and is analogous to the resonance broadening familiar from wave-particle
interactions in plasmas \markcite{Dupree72}({Dupree} 1972). This effect is small for
nearly parallel propagation of Alfven waves, since $V_{g}\approx \const $.
However, for a more realistic spectrum composed of a mixture of Alfven
and magnetosonic waves with a sizable off-axis component, the effect
is significant. The second flavor is simply direct scattering of the
Alfven quanta in $k$, regardless of $V_{g}$ variations. This contributes
to the resonance broadening as $\Delta \omega \sim D_{k}/\Delta k^{2}$,
where $\Delta k$ is the width of the wave packet. The first effect
is usually larger in most situations when the field of scatterers
is weak ($\rho _{q}\ll \rho _{0}$) but in our case, rather the opposite
is true \markcite{KangJR92}({Kang} {et~al.} 1992), so that the second effect turns out to
be stronger. This is ultimately due to the weak dispersion of quasiparallel
Alfven waves.

For steepened acoustic fluctuations, we can assume the scatterer field
is simply an ensemble of shock-like discontinuities, so

\begin{equation}
\frac{\rho _{q}^{2}}{\rho _{0}^{2}}\simeq A\left(\frac{q_{0}}{q}\right)^{2}\label{rho_{q}}\end{equation}
This form is appropriate for a shock ensemble with a characteristic
amplitude $\sim \sqrt{A}\sim \Delta \rho /\rho _{0}$ and a separation
$\sim 1/q_{0}$. This kind of density perturbation structure is to
be expected from the nonlinearly developed Drury instability \markcite{KangJR92}({Kang} {et~al.} 1992)
since that would produce a pattern of discontinuities. From equation
(\ref{Dk2}) we then obtain

\begin{equation}
D_{k}\simeq \frac{3}{2}Aq_{0}\frac{k^{2}V_{A}^{2}}{V_{G}}\left(\frac{\pi }{2}-\frac{1}{y}\right)\label{Dk3}\end{equation}
 where $y^{3}=D_{k}V_{g}^{\prime 2}/q_{0}V_{g}^{3}$. Further progress
is possible only for the limiting cases of weak and strong scattering.
Therefore, in the case of strong scattering ($y\gg 1$) we have

\begin{equation}
D_{k}\approx \frac{3\pi }{4}Aq_{0}\frac{V_{A}^{2}}{V_{g}}k^{2}\equiv \nu k^{2}\label{Dk4}\end{equation}
where $V_{g}^{\prime }$ cancelled out. In the opposite case of weak
scattering ($y\ll 1$) we obtain

\begin{equation}
D_{k}\approx \left(\frac{3}{2}A\right)^{3/2}\frac{q_{0}k^{3}V_{A}^{3}V_{g}^{\prime }}{V_{g}^{3}}\label{Dk5}\end{equation}
Restricting consideration to the case of strong scattering with $D_{k}$
as given by equation (\ref{Dk4}), the steady state limit of equation
(\ref{wkeI}) becomes

\begin{equation}
\frac{\partial }{\partial k}k^{2}\frac{\partial I}{\partial k}-\frac{u_{1}}{\nu }\frac{\partial I}{\partial x}=-\left.2M_{A}\frac{u_{1}}{\nu k}\frac{\partial P}{\partial x}\right|_{pk=m\omega _{c}}\label{wkeI2}\end{equation}
In the case of weak diffusion, with no spectral spreading (\ie $\nu \to 0$),
the turbulence level is simply directly proportional to the particle
partial pressure

\begin{equation}
I\simeq 2M_{A}k^{-1}P\left(p=\omega _{c}/kc\right)\label{Ismallbet}\end{equation}
where $p$ is normalized to $mc$. It is convenient to introduce the
following dimensionless variables

\[
\tau =\frac{\nu x}{u_{1}},\; \; Q=2M_{A}P/k,\; \; v=\ln \left(kc/\omega _{c}\right)\; \and \; \Psi =Ie^{\frac{v}{2}+\frac{\tau }{4}}\]
so that the equation for $\Psi $ becomes

\[
\frac{\partial \Psi }{\partial \tau }-\frac{\partial ^{2}\Psi }{\partial v^{2}}=e^{\frac{v}{2}+\frac{\tau }{4}}\frac{\partial Q}{\partial \tau }\]
This can be solved to obtain the wave spectral density $I$

\begin{equation}
I=\int _{-\infty }^{\tau }\frac{d\tau ^{\prime }e^{-\frac{1}{4}\left(\tau -\tau ^{\prime }\right)}}{\sqrt{4\pi \left(\tau -\tau ^{\prime }\right)}}\int _{-\infty }^{+\infty }dv^{\prime }e^{-\frac{\left(v-v^{\prime }\right)^{2}}{4\left(\tau -\tau ^{\prime }\right)}+\frac{1}{2}\left(v-v^{\prime }\right)}\frac{\partial Q}{\partial \tau ^{\prime }}\left(v^{\prime }\right)\label{I:sol}\end{equation}
By introducing the dimensionless coordinate $\xi =x/L$, it easy to
see that the main parameter that determines the behaviour of the solution
for $I$ is the stregth parameter $S$

\[
S=\frac{\nu L}{u_{1}}=\frac{t_{conv}}{t_{s}}\]
where $t_{conv}=L/u_{1}$ is the precursor (of length $L$) crossing
time and $t_{s}=1/\nu $ is the refractive scattering time for the
wave. In the limit $S\to 0$ one simply obtains $I=Q$, equation (\ref{I:sol}),
which is equivalent to equation (\ref{Ismallbet}) above. In the more
interesting case $S\gg 1$, by using the steepest descent method we
obtain

\begin{equation}
I=\frac{1}{S}\int _{-\infty }^{+\infty }dv^{\prime }e^{-\left(v-v^{\prime }\right)H\left(v-v^{\prime }\right)}\frac{\partial }{\partial \xi ^{\prime }}Q\left(\xi ^{\prime },v^{\prime }\right)_{\xi ^{\prime }=\xi -\frac{1}{\beta }\left|v^{\prime }-v\right|}\label{Ibigbet}\end{equation}
where $H$ is the Heavyside function. One sees that there is a $S\gg 1$
reduction in the wave spectral density as compared to the scattering
free case given by equation (\ref{Ismallbet}). To clarify this last
result, let us specify the particle distribution function at some
dimensionless distance $\xi $ ahead of the shock. We assume that
the upstream medium is at $x<0$ and since we neglected the flow compression
in equation (\ref{wkeI2}) we assume the stationary solution of the
diffusion-convection equation, equation (\ref{dc1}) has the following
form

\begin{equation}
f\left(p,x\right)=f_{0}\left(p\right)\exp \left(\frac{u_{1}x}{\kappa }\right)\label{dc:sol}\end{equation}
Note that in the case of a modified shock precursor, the solution
has a similar form, except for the modified term in the exponent.
That is replaced by an integral over $x$ and a multiplicative factor
of the order of unity, depending slowly on $p$ \markcite{m97a}({Malkov} 1997). For
this simple estimate we can assume that $\kappa $ is a linear function
of $p$, \ie 

\begin{equation}
\kappa \simeq \kappa _{b}\frac{p}{p_{b}}\label{kappab}\end{equation}
where we have introduced the lower cut-off momentum $p_{b}$ from
equation (\ref{dc:sol}) in an obvious manner, $\kappa _{b}\left(x\right)\equiv \kappa \left[p_{b}\left(x\right)\right]=\left|x\right|u_{1}$.
We also assume that the spectrum has an upper cut-off at $p_{max}$
and $\kappa $ scales linearly with momentum up to $p_{max}$, so
that $\kappa _{max}\simeq \kappa _{b}\left(p_{max}/p_{b}\right)$.
We account for the fact that $L=\kappa _{max}/u_{1}$. 

From the point of view of confinement and thus acceleration improvement
due to spectral transfer it is important to understand whether the
spectrum that is resonantly generated by a group of already accelerated
particles near $p_{max}$ propagates to lower $k$, so as to facilitate
confinement and acceleration of particles with $p>p_{max}$. Note
that the wave compression due to the nonlinear shock modification
in the CR precursor leads to an opposite evolution in $k$. 

Since the waves are originally driven by accelerated particles, it
is instructive to express the wave energy density through the particle
pressure. Note that there is a coordinate dependent low energy cut-off
$p_{b}\left(x\right)$ on the particle spectrum which physically means
that low-energy particles cannot diffuse far ahead of the subshock
(see equation {[}\ref{dc:sol}{]}). Normalizing the particle pressure
to the shock ram pressure it is convenient to express the former one
as an acceleration efficiency

\begin{equation}
\epsilon \left(x\right)\equiv \frac{P_{c}\left(x\right)}{\rho _{1}u_{1}^{2}}=\frac{4\pi }{3}\frac{mc^{2}}{\rho _{1}u_{1}^{2}}\int _{p_{b}\left(x\right)}^{p_{max}}dpp^{3}f_{0}\left(p\right)\label{effic}\end{equation}
where we have ignored the exponential factor in the distribution function,
equation (\ref{dc:sol}) since the integral in $p$ effectively runs
above $p_{b}\left(x\right)$. Assuming then that $p_{b}$ is reasonably
close to $p_{max}$, $p_{b}\la p_{max}$, which means that $x$ is
taken sufficiently far from the sub-shock, but not so far so there
are not enough particles or waves. This case is therefore interesting,
in that there is still sufficient space for spectrum spreading before
waves are convected into the subshock. In this limit, we can estimate
the spectrum from equation (\ref{Ibigbet}), expressing it for convenience
in terms of the resonant $p$, rather than in terms of $k$, as: 

\begin{equation}
I(x)=\frac{2M_{A}}{\beta }\epsilon \left(x\right)\left\{ \begin{array}{cc}
 1, & p>p_{max}\\
 p/p_{max}, & p<p_{max}\sim p_{b}\end{array}
\right.\label{eq:Iofx}\end{equation}
The connection of this result to scattering of the Alfven quanta in
$k$ can be seen directly from equation (\ref{wkeI2}). It corresponds
to the solution of equation (\ref{wkeI2}) in two regions of wave
number $k$, where the source is absent, $P\simeq 0$. The low-$k$
asymptotics (high $p$) is simply a solution with constant $I\left(k\right)=\const $,
whereas the high-$k$ part of the solution corresponds to the constant
flux of the wave density in $k$ space, $k^{2}\partial I/\partial k=\const $.
These limiting forms obviously match in the region where waves are
generated, \ie where $p\sim p_{b}\sim p_{max}$. \emph{Obviously,
equation} (\ref{eq:Iofx}) \emph{indicates that diffusive refraction
in $k$ space leads to the confinement and acceleration of higher
energy particles, with} $p>p_{max}$.

At this point, it is appropriate to discuss, more generally, the flow
of wave population density $N$ and wave energy density $\mathcal{E}$
in $k$ space which results from the stochastic scattering of Alfven
waves off precursor density fluctuations which we described above.
Since this interaction is modeled (at the quasilinear level) by diffusion
in $k$ space, we can write the relevant part of the wave population
kinetic equation as

\[
\frac{\partial \left\langle N\right\rangle }{\partial t}=\frac{\partial }{\partial k}D_{k}\frac{\partial \left\langle N\right\rangle }{\partial k}+\ldots \]

so a gradient in the population $\left\langle N\right\rangle $ produces
a flux in $k$

\[
\Gamma _{k}=-D_{k}\frac{\partial \left\langle N\right\rangle }{\partial k}\]
Thus, since waves are excited at $kr_{g}\sim 1$, $\partial \left\langle N\right\rangle /\partial k>0$
for $kr_{g}<1$ and $\partial \left\langle N\right\rangle /\partial k<0$
for $kr_{g}\gg 1$. Thus, $\Gamma _{k}<0$ for $kr_{g}<1$, suggestive
of a flux of the Alfven wave population density to \emph{large} scales.
The total wave population

\[
\mathcal{N}=\int _{k_{min}}^{k_{max}}dk\left\langle N\right\rangle \]
evolves according to 

\[
\frac{d\mathcal{N}}{dt}=D_{k}\left.\frac{\partial \left\langle N\right\rangle }{\partial k}\right|_{k_{min}}^{k_{max}}\]
and thus is determined by the slope of the spectrum $N$ in the high
and low $k$ limits since the wave energy density $\mathcal{E}=\omega _{k}N$,
the evolution of the total energy 

\[
E_{\omega }=\int _{k_{min}}^{k_{max}}dk\mathcal{E}\left(k\right)\]
is given by

\begin{eqnarray*}
\frac{dE}{dt} & = & \int _{k_{min}}^{k_{max}}dk\omega _{k}\frac{\partial }{\partial k}D\frac{\partial \left\langle N\right\rangle }{\partial k}\\
= & \omega _{k} & D_{k}\left.\frac{\partial \left\langle N\right\rangle }{\partial k}\right|_{k_{min}}^{k_{max}}-\int _{k_{min}}^{k_{max}}dk\left(\frac{\partial \omega }{\partial k}\right)D_{k}\frac{\partial \left\langle N\right\rangle }{\partial k}
\end{eqnarray*}
A detailed spectral evolution calculation, beyond the scope of this
paper, is required to precisely determine the sign of $dE/dt$. That
said, it seems very likely indeed that the \emph{wave number integrated}
contribution (second term on r.h.s.) is both dominant, and is \emph{negative}.
This follows from $D_{k}>0$, $V_{g}=\partial \omega /\partial k>0$
(\ie the counter-propagating streams do not precisely balance) and
$\partial \left\langle N\right\rangle /\partial k\ge 0$, since there
is a greater population density at small scales. Thus $dE/dt<0$,
which is also consistent with the outcome of the modulational instability
calculation. Since waves are scattered by refraction by precursor
density fluctuation, we know that the nonlinear interaction leaves

\[
\frac{d}{dt}\left(E_{w}+E_{\rho }\right)=0\]
so $dE_{\rho }/dt>0$, \ie the \emph{decay instability of short wavelength}
($kr_{g}\sim 1$) \emph{Alfven waves amplifies the precursor density
perturbation energy and depletes Alfven wave energy while it scatters
Alfven wave energy towards larger scales.}

The upshot of this decay/modulational interaction process can thus
be viewed as an energy transfer process in $k$ space which increases
the population of high energy cosmic rays which are confined to the
shock and precursor. This may be seen by considering the sequence
below, \ie

i.) as usual, energetic CRs generate Alfven waves with $kr_{g}\sim 1$

ii.) these waves scatter off of ambient density perturbations in the
precursor via decay instability, thus:

a.) producing larger wavelength Alfven waves

b.) amplifying the density perturbation field and so producing a flow
of fluctuation energy (including magnetic energy) toward longer scales.
At the same time:

a.) the longer wavelength Alfven waves will confine higher energy
particles

b.) the precursor density perturbations will produce high efficiency
acceleration (\ie beyond the Bohm limit) of particles with $p\ge p_{*}$
(where $p_{*}$ corresponds to the {}``knee'') by scattering these
particles in a converging flow. This process does \emph{not} require
particles to cross and re-cross the shock, itself.

In this way, we see a \emph{direct connection between fluctuation
energy flow to larger scales and achieving increased particles energy}
during the acceleration process. Note also that two acceleration processes
(\ie the traditional one involving multiple shock crossing via confinement
by turbulent pitch angle scattering and the recently proposed one
involving scattering off inhomogeneities in the converging percursor
flow) can work together by exploiting the wave scattering mechanism
discussed here.

\section{Conclusions and Discussion\label{sec:Conclusions-and-Discussion}}

In this paper, we have examined the dynamics and generation of mesoscale
magnetic field in diffusive shock acceleration. The principal results
of this paper are:

\begin{enumerate}
\item the identification of a new (in the context of DSA theory) nonlinear,
multiscale interaction mechanism involving precursor density fluctuations
and resonantly generated Alfven waves was identified. This mechanism
is one of modulational or decay instability of Alfven waves in a scattering
field of density perturbations.
\item the demonstration that such interaction: \begin{enumerate} 
\item generates larger scale waves, which in turn confine higher energy
particles to the shock, thus allowing their acceleration
\item can also enhance the level of density fluctuations in the precursor,
and so assist in the (recently proposed) high efficiency acceleration
by scattering cosmic ray particles off inhomogeneities in the converging
flow.\end{enumerate}
\item the explicit construction of a strong turbulence theory, based on
modulational interaction, for \emph{calculating} the spreading of
Alfven wave energy. The principal mechanism by which Alfven wave energy
is scattered to larger scales is shown to be random refraction by
the spectrum $\left|\left(\nabla \widetilde{\rho }/\rho \right)_{q}\right|^{2}$
- of density gradient fluctuations in the precursor flow. These density
fluctuations can be produced by the modulational process itself, or
develop from precursor instabilities, such as the Drury instability.
\item the identification of the critical parameter which governs this $k$
space diffusion process, namely $S=L/u_{1}\tau _{s}$, where $1/\tau _{s}\sim D_{k}/\Delta k^{2}$
is the refractive scattering time.
\item the demonstration that for $S>1$, diffusion in $k$ will generally
broaden and suppress narrow band spectra at $kr_{g}\sim 1$, while
scattering energy to longer (and shorter) wavelengths.
\end{enumerate}
The relation of this type of mechanism to existing concepts in DSA
theory was discussed thoroughly. 

The results of this paper have several interesting implications for
high energy cosmic ray acceleration. First, given that density fluctuations
are always present via the Drury instability, waves at $kr_{g}\sim 1$
will not grow to large amplitude ($\delta B\ga B_{0}$), but rather
will have their energy diffused in $k$ to a broad band of larger
and smaller scales. Note that this new nonlinear mechanism of saturation
of the resonant RMS magnetic energy is independent of the quasilinear-type
instability saturation mechanisms discussed in the Introduction, which
were considered earlier by \eg \markcite{McKVlk82}{MacKenzie} \& {Voelk} (1982) and \markcite{AchtBland86}{Achterberg} \& {Blandford} (1986).
Thus, theories which predict the generation of strong, small scale
fields without considering scattering by precursor fluctuations probably
have significantly \emph{overestimated} the strength of the field
at $kr_{g}\sim 1$. Second, the modulational mechanism presented here,
while not a {}``dynamo'', in a strict sense, is a robust and universal
means to scatter wave energy to larger scales and so confine higher
energy particles (\ie $p>p_{max}$). Hence, it constitutes a novel
means for enhanced acceleration. A quantitative calculation of the
resulting energy spectrum clearly requires solution of the coupled
kinetic equations for Alfven waves, acoustic perturbations, along
with the evolution equation for the energetic particle distribution
function $f$, and the nonlinear shock conditions. This is obviously
a formidable task, and one which will require significant effort in
the future. However, a more tractable approach \ie a first step is
to solve only the coupled equations for the resonantly generated Alfven
wave population and the energetic particle distribution function,
incorporating diffusion in wavenumber by a prescribed spectrum of
ambient density fluctuations associated with Drury instability. This
calculation, which is a fairly modest extension of the analysis in
Sec.\ref{sub:Estimates-of-Alfven} of this paper, should reveal the
quantitative relations between the precursor scatterer field spectrum
and the energy spectrum of accelerated particles confined to the shock
by turbulent pitch angle scattering. The results of this calculation
will be presented in a future publication. However, the reader should
also keep in mind that the modulational instability will amplify any
perturbations in the precursor, thereby also enhancing acceleration
by scattering of CRs by inhomogeneities in its converging flow. Thus,
it increasingly seems that the most energetic particles result as
much from precursor dynamics, as from the traditionally invoked crossing
of the subshock discontinuity, per se.

\acknowledgements{We thank L. Drury, H. Li and M. Medvedev for stimulating discussions.
This work was supported by NASA under grant ATP03-0059-0034 and by
the U.S. DOE under Grant No. FG03-88ER53275}

\bibliography{}

\begin{figure}
\epsscale{1.0}
\plotone{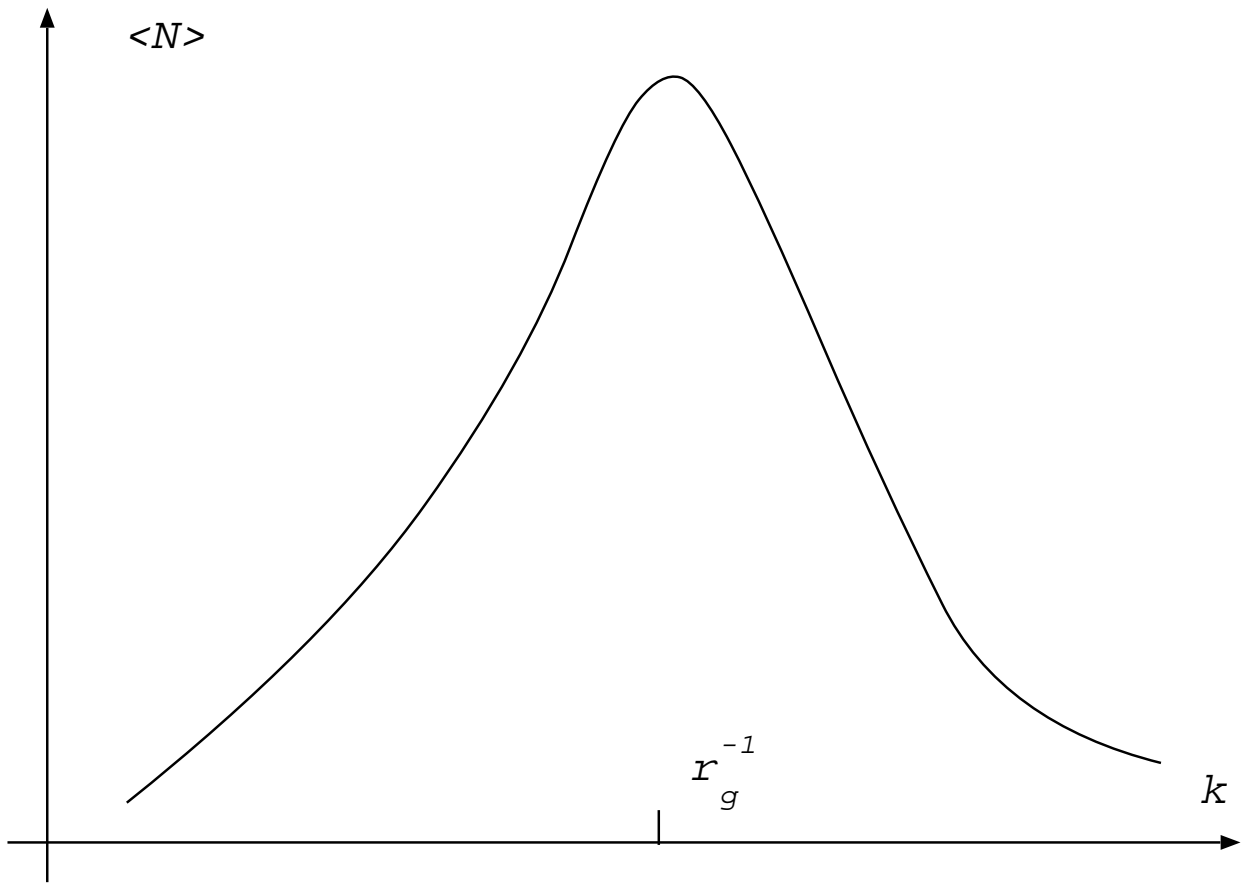}

\caption{Formation of inverted population of Alfven quanta under the condition
of localized (at $kr_{g}\sim 1$) driver.\label{cap:Formation-of-inverted}}
\end{figure}

\end{document}